\documentclass[%
reprint,
 superscriptaddress,
 preprintnumbers,
 nofootinbib,
 amsmath,amssymb, 
 aps,
 prl,
longbibliography
]{revtex4-1}

\usepackage[utf8]{inputenc}

\usepackage{csquotes}

\usepackage[dvipsnames]{xcolor}
\usepackage{xspace}
\usepackage{graphicx}
\usepackage{cancel} 
\usepackage{multirow}
\usepackage{array, booktabs, makecell, multirow}
\setcellgapes{5pt}
\graphicspath{{./figures/}}
\usepackage{subcaption} 
\usepackage{textcomp}
\usepackage{slashed}

\usepackage{amsmath}
\usepackage{mathtools}
\usepackage{bm} 

\usepackage{array} 
\usepackage{booktabs} 
\usepackage[skip=0.5\baselineskip]{caption} 
\usepackage[colorlinks=true]{hyperref}
\hypersetup{
  linkcolor=NavyBlue,
  citecolor=NavyBlue,
  filecolor=Red,
  urlcolor=NavyBlue,
  menucolor=Red,
  runcolor=Red
}

\usepackage{acronym}
\acrodef{UV}{ultraviolet}
\acrodef{IR}{infrared}

\usepackage{cleveref} 

\newcommand{\moller}{M{\o}ller\xspace}

\newcommand{\eps}{\epsilon}

\newcommand{\beq}{\begin{eqnarray}}
\newcommand{\eeq}{\end{eqnarray}}
\newcommand{\rd}{\mathrm{d}}
\newcommand{\rJ}{\mathrm{J}}
\newcommand{\rT}{\mathrm{T}}

\newcommand{\rI}{\mathrm{I}}
\newcommand{\cA}{\mathcal{A}}
\newcommand{\cC}{\mathcal{C}}

\newcommand{\rs}{y}
\newcommand{\rt}{z}

\DeclareMathOperator{\EK}{K}

\newcommand{\dfS}{s}
\newcommand{\dfT}{t}
\newcommand{\dfU}{u}

\begin{document}

\preprint{BONN-TH-2023-13, TUM-HEP-1479/23}

\title{Two-loop QED corrections to the scattering of four massive leptons}

\author{Maximilian Delto}
\affiliation{Physik Department, James-Franck-Straße 1, Technische Universität München, D--85748 Garching, Germany}

\author{Claude Duhr}
\affiliation{Bethe Center for Theoretical Physics, Universität Bonn, D-53115, Germany}

\author{Lorenzo Tancredi}
\affiliation{Physik Department, James-Franck-Straße 1, Technische Universität München, D--85748 Garching, Germany}

\author{Yu Jiao Zhu}
\affiliation{Bethe Center for Theoretical Physics, Universität Bonn, D-53115, Germany}

\date{\today}

\preprint{}

\begin{abstract}
We study two-loop corrections to the scattering amplitude of four massive leptons in quantum electrodynamics. 
These amplitudes involve previously unknown elliptic Feynman integrals,
which we compute analytically using the differential equation method.
In doing so, we uncover the details of the
elliptic geometry underlying this scattering amplitude and show how to
exploit  its properties to obtain compact, easy-to-evaluate 
series expansions that describe the scattering of four massive leptons in QED 
in the kinematical regions relevant for Bhabha and \moller scattering processes.
\end{abstract}

\maketitle


In recent years, particle physics has seen several interesting developments in 
experiments at the low-energy precision frontier. 
Among these are the discrepancy between theory predictions and the experimental value for the muon anomalous magnetic moment, most recently measured to $0.20~\text{ppm}$ at Fermilab \cite{Muong-2:2023cdq}, 
as well as the so called `proton radius puzzle'.
 The latter consists in a discrepancy between the proton charge radius 
 as determined in~\cite{Pohl:2010zza,Antognini:2013txn} compared to previous results~\cite{Mohr:2008fa}. The upcoming PRad II experiment \cite{PRad:2020oor} will perform an independent measurement to 
 attempt to resolve this inconsistency.

This experimental program requires matching efforts on the theoretical side to provide equally
precise and reliable predictions. An important part of these efforts is the recent development of the Monte Carlo event generator \texttt{McMule} \cite{Banerjee:2020rww}. With the newly developed next-to-soft stabilization~\cite{Banerjee:2021qvi,Banerjee:2021mty} for real-virtual corrections, \texttt{McMule} has the potential to describe Bhabha \cite{Bhabha:1936zz} ($e^+e^- \to e^+e^-$)  and \moller scattering ($e^-e^- \to e^-e^-$) at the fully differential level up to next-to-next-to-leading order (NNLO) in Quantum Electrodynamics (QED). 
While Bhabha scattering is important for luminosity measurements at lepton colliders, \moller scattering is the main source of systematic uncertainty for the PRad II experiment~\cite{PRad:2020oor} quoted above. \moller scattering is also important in searches for parity violation and for precise measurements of the weak mixing angle~\cite{MOLLER:2014iki}. Finally, precision measurements
of \moller scattering at very low energies ($2.5\,\text{MeV}$)~\cite{Epstein:2019gjd} have 
also been undertaken recently.

The remaining outstanding ingredient to make theoretical studies 
in NNLO QED at arbitrary energy scales
possible, are the two-loop virtual amplitudes for the scattering of four massive leptons,
retaining full dependence on the lepton mass.
The calculation of these virtual corrections has received much attention in the last decades. 
In QED with massless leptons, these amplitudes were computed more than two decades ago~\cite{Bern:2000ie}. Full event simulations at leading and next-to-leading order (NLO)~\cite{Jadach:1996gu,Montagna:1998sp,Montagna:1998kp},
as well as power-suppressed mass effects up to NNLO have also been studied in detail~\cite{Penin:2005kf,Penin:2005eh,Mitov:2006xs,Becher:2007cu,Actis:2007gi,Penin:2016wiw}, while
fermionic loop corrections were computed with full mass dependence in~\cite{Bonciani:2004gi}. 
Moreover, also logarithmically enhanced electroweak contributions have been considered to NNLO~\cite{Kuhn:2001hz,Feucht:2004rp,Jantzen:2005az,Penin:2011aa}. 
However, even though the computation of the relevant two-loop integrals was initiated already more than two decades ago~\cite{Smirnov:2001cm,Heinrich:2004iq,Czakon:2004wm,Czakon:2005gi,Czakon:2005jd,Czakon:2006hb,Czakon:2006pa,Duhr:2021fhk,Henn:2013woa}, complete results for the two-loop virtual amplitudes where the full mass-dependence on the lepton mass is retained, are still not available, mostly due to the complexity of the integrals involved.  

In this letter, we move an important step towards the exact inclusion of mass effects
to Bhabha and \moller scattering up to NNLO in QED, by performing the first fully massive calculation of the two-loop QED 
corrections to the polarized and unpolarized scattering amplitude of four massive leptons.
While retaining the full mass dependence renders the required two-loop integrals 
considerably more complicated, it will allow us to study the phenomenological 
impact of so-far neglected mass-effects in extreme regions of phase-space in upcoming 
phenomenological studies. In addition to their phenomenological relevance, 
these amplitudes also provide us with an invaluable playground to test 
recent developments in the theory of elliptic generalizations of multiple polylogarithms~\cite{Broedel:2014vla,Adams:2016xah,Adams:2017ejb,Remiddi:2017har,Broedel:2018qkq,Duhr:2019rrs,Walden:2020odh} and about the generalization of
so-called canonical differential equations~\cite{Henn:2013pwa} to arbitrary geometries~\cite{Primo:2016ebd,Primo:2017ipr,Frellesvig:2021hkr,Giroux:2022wav,Dlapa:2022wdu,Pogel:2022vat,Frellesvig:2023iwr,Gorges:2023zgv,Jiang:2023jmk}.


\section{Kinematics and tensor decomposition}
\label{sec:notation-tensor-decomposition}
We work in QED with one single type of massive lepton, which for definiteness we refer to as the electron. We study higher-order corrections to the scattering of four electrons
\begin{align}
    0 \to e^+(p_1) + e^-(p_2) + e^-(p_3)  +e^+(p_4) \,,
\end{align}
where all momenta are outgoing and satisfy the on-shell condition $p_i^2=m^2$, $i=1\dots4$, as well as momentum conservation $p_1^\mu+p_2^\mu+p_3^\mu+p_4^\mu=0$.
The corresponding amplitude, $\mathcal{A}(1_{e^+},2_{e^-},3_{e^+},4_{e^-})$, can be parameterized as a function of the fermion mass $m$, and three Mandelstam invariants
\begin{align}
    \dfS =(p_1+p_2)^2\,, ~~ \dfT=(p_1+p_3)^2\,,~~ \dfU=(p_2+p_3)^2\,,
\end{align}
where, due to momentum conservation, $\dfS+\dfT+\dfU=4m^2$.

Following~\cite{Peraro:2019cjj,Peraro:2020sfm}, we work in 't Hooft-Veltman dimensional
regularization scheme~\cite{tHooft:1972tcz} (tHV) and
decompose the scattering amplitude into eight independent Lorentz-covariant, \textit{physical} tensors  $T_i$ and respective scalar form factors $\mathcal{F}_i$,
\begin{align}
\label{eqn:form_factor_decomp}
     \mathcal{A}(1_{e^+},2_{e^-},3_{e^-},4_{e^+}) = \sum_{i=1}^{8} \mathcal{F}_i \, T_i \,.
\end{align}
We choose the tensors as
\begin{align}
\label{eqn:definition_tensors_T}
T_{1} ={} & m^2 \times t_1 \,, & T_{2} ={} &m \times \left[  t_2 + t_3 \right] \nonumber \\
T_{3} ={} & t_4 \,, & T_{4} ={} & m^2 \times t_5 \,, \nonumber \\
T_{5} ={} & m \times \left[ t_6 + t_7 \right] + t_8 \,, & T_{6} ={} & m \times \left[ t_6 + t_7 \right] - t_8 \nonumber \\
T_{7} ={} & m \times \left[ t_2 - t_3 \right] \,, & T_{8} ={} & m \times \left[ t_6 - t_7 \right] \,, 
\end{align}
where the spinor chains $t_i$ are defined as
\begin{align}
\label{eqn:definition_tensors_t}
    t_i = \overline{U}_e(p_2) \, \Gamma^{(1)}_i \, V_e(p_1)   \times  \overline{U}_e(p_3) \, \, \Gamma^{(2)}_i \, V_e(p_4) \,,
\end{align}
and $\mathbf{\Gamma}_i=\{\Gamma^{(1)}_i,\Gamma^{(2)}_i\}$ represent the following sets of Dirac matrices
\begin{align}
\label{eqn:definition_tensostructures}
  \mathbf{\Gamma}_1={}&\{1,1\}\,,&\mathbf{\Gamma}_2={}&\{\slashed{p}_3,1\}\,, \nonumber \\
  \mathbf{\Gamma}_3={}&\{1,\slashed{p}_2\}\,,&\mathbf{\Gamma}_4={}&\{\slashed{p}_3,\slashed{p}_2\}\,, \nonumber \\
  \mathbf{\Gamma}_5={}&\{\gamma^{\mu_1},\gamma_{\mu_1}\}\,,&\mathbf{\Gamma}_6={}&\{\slashed{p}_3\gamma^{\mu_1},\gamma_{\mu_1}\}\,, \nonumber \\
  \mathbf{\Gamma}_7={}&\{\gamma^{\mu_1},\slashed{p}_2\gamma_{\mu_1}\}\,,&\mathbf{\Gamma}_8={}&\{\slashed{p}_3\gamma^{\mu_1},\slashed{p}_2\gamma_{\mu_1}\} \,.\,
\end{align}
In this computational scheme, external momenta and polarizations are considered four-dimensional, 
while internal states and loop momenta are treated in $D$ dimensions. One can then show that the number of tensors is equal to the number of independent chirality configurations to all orders in perturbation theory~\cite{Peraro:2019cjj,Peraro:2020sfm}.
In our case there are $2^4=16$ configurations, of which only half are independent in 
a parity-invariant theory such as QED, which matches the eight tensors above. Furthermore, 
we note that the process under consideration is invariant under the simultaneous 
exchange $p_2\leftrightarrow p_3$ and $p_1\leftrightarrow p_4$. We find that
under this transformation two tensor structures
are mapped onto each other, i.e. $t_2\leftrightarrow t_3$ and  $t_6\leftrightarrow t_7$, cf.~(\ref{eqn:definition_tensostructures}), which in turn implies that $T_{7,8}$ are odd. Accordingly, by symmetry we conclude that the corresponding form factors must be zero $\mathcal{F}_7= \mathcal{F}_8=0$
to all orders.

Following the standard approach, we compute the form factors in~(\ref{eqn:form_factor_decomp}), 
by defining a set of projection operators $ \mathcal{P}_i = \left[ \left( T^{\dagger} \cdot T \right)^{-1} \right]_{ik} T^{\dagger}_k$ as combinations of \emph{dual tensors} $T_i^\dagger$.
Here ``$\cdot$" denotes the scalar product between tensors and their dual, 
which is realized in practice by summation over spins of the external fermions,
such that $\mathcal{F}_i = \mathcal{P}_i \cdot \mathcal{A}$.

By applying the projectors on the corresponding relevant QED Feynman diagrams, 
we can express each form factor as a linear combination of scalar
Feynman integrals and organize the one- and two-loop integrals
into several integral topologies.
On the technical level, our computation proceeds as follows. We begin by generating relevant Feynman diagrams with \texttt{QGRAF} \cite{Nogueira:1991ex}. Using the computer algebra system \texttt{FORM}~\cite{Vermaseren:2000nd,Kuipers:2012rf,Kuipers:2013pba,Ruijl:2017dtg}, we insert Feynman rules and apply projectors. We employ the public tool \texttt{Reduze2}~\cite{vonManteuffel:2012np} to find mappings onto 
topologies and to expose their symmetries. Finally, with the help of \texttt{Kira}~\cite{Maierhofer:2017gsa,Maierhofer:2018gpa,Klappert:2020nbg} we solve the required integration-by-parts (IBP) relations \cite{Tkachov:1981wb,Chetyrkin:1981qh} and reduce all integrals to 267 master integrals. 
This is achieved following Laporta's algorithm \cite{Laporta:2000dsw}, improved by finite field techniques \cite{vonManteuffel:2014ixa,Peraro:2016wsq}.


\section{Canonical bases for the  non-planar Feynman integrals}
\begin{figure}[th]
	\begin{minipage}[t]{.228\textwidth}
		\includegraphics[width=\textwidth]{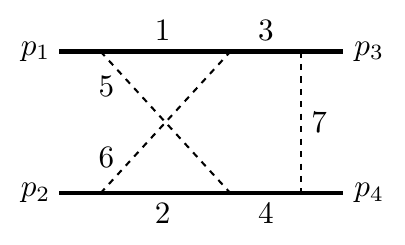}\\	
	\end{minipage}
		\quad\quad
	\begin{minipage}[t]{.21\textwidth}
		\includegraphics[width=1\textwidth]{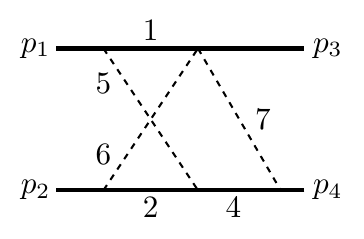}\\
	\end{minipage}
\caption{
The non-planar topology (left) and its next-to-top sector (right) with 6 master integrals.
Solid lines correspond to massive propagators of mass $m$, dashed lines correspond to massless propagators.
}
		\label{fig-non-planar-top}
\end{figure}

While all planar two-loop topologies have been known in fully analytic form for some time~\cite{Henn:2013woa,Duhr:2021fhk},
their non-planar counterparts have remained elusive due to the appearance of new mathematical functions of elliptic type. 
In particular, we are interested in the non-planar family of integrals displayed 
in the left graph of fig~(\ref{fig-non-planar-top}).
%
We work in dimensional regularization and define 
the integrals as
\beq
\label{def_integral}
\lefteqn{
 \rI_{a_1 a_2 a_3 a_4 a_5 a_6 a_7 a_8 a_9}\left( D, \frac{\dfS}{m^2}, \frac{\dfT}{m^2} \right)
 = } & &
 \nonumber \\ 
 & &
 e^{2 \gamma_E \eps}
 \left(\mu^2\right)^{\sum\limits_{j=1}^9 a_j - D}
 \int \frac{\rd^Dk_1}{i \pi^{\frac{D}{2}}} \frac{\rd^D k_2}{i \pi^{\frac{D}{2}}}
 \prod\limits_{j=1}^9 \frac{1}{ P_j^{a_j} },
\eeq
where $\gamma_E$ denotes the Euler-Mascheroni constant, 
$D=4-2\eps$ is the dimension of space-time,
and  $\mu$ is an auxiliary scale introduced to render Feynman integrals dimensionless.
The propagators are given by
\begin{align}
 P_1 & =k_1^2 - m^2,
 &
 P_2 & = (k_1 - k_2 - p_2)^2 - m^2,
 \nonumber \\
 P_3 & = k_2^2 -  m^2,
 &
 P_4 & = (k_2 + p_1 + p_2)^2 - m^2,
 \nonumber \\
 P_5 & = (k_1 + p_1)^2,
 &
 P_6 & = (k_1 - k_2)^2,
 \quad
  P_7  = (k_2 - p_3)^2,
 \nonumber \\
  P_8 & = (k_2 + p_1)^2,
  &
 P_9 & = (k_1 - p_3)^2.
 &
\end{align}
The integrals in~\eqref{def_integral} are  functions of  
homogeneous coordinates $[\,\dfS\,:\,\dfT\,:\,m^2\,]$ on $\mathbb{CP}^2$ and,
without loss of generality, we may set $\mu=m$, or equivalently work on the patch $[\,\rs\,:\,\rt\,:\,1\,]$ with $\rs=\dfS/m^2$ and $\rt=\dfT/m^2$. For definiteness, we will display our formulas in the region 
$
  \dfS > 4m^2 \,,   \dfT < 0\,,
$
though all results can also be easily continued to any other kinematic region.
By solving IBP identities, all integrals 
can be expressed in terms of $52$ independent master integrals.
The latter fulfil a system of first-order partial-differential equations~\cite{Kotikov:1990kg,Kotikov:1991hm,Kotikov:1991pm,Remiddi:1997ny,Gehrmann:1999as} 
in the kinematical invariants
 \beq
 \rd \vec{\rI} & = & A\left(\eps,\rs,\rt\right) \vec{\rI}\,.
\eeq
To solve this system, it is useful to search for a basis transformation to a so-called \emph{$\eps$-factorized form}:
 \beq
 \label{eq:eps-form}
 \rd \vec{ \rJ} & = &\eps \mathrm{A} \left(\rs,\rt\right) \vec{ \rJ}, \quad \vec{ \rJ}=\mathrm{U}(\rs,\rt,\eps) \vec{\rI}\,.
\eeq
Such a system  can be formally solved
 by a path-ordered exponential 
 \beq
 \label{eq:g_path_sol}
\vec{\rJ} (\rs,\rt,\eps)= \mathbb{P}\textrm{exp}\left[\eps\int_{\gamma} \mathrm{A}\right]\,\vec{\rJ}_0(\eps,\rs_0,\rt_0)\,,
\eeq
where the path $\gamma$ connects  the initial boundary point  $(\rs_0,\rt_0)$ to a generic point $(\rs,\rt)$.
In the polylogarithmic case, if the matrix $ \mathrm{A}$ can be expressed only
through logarithmic differential forms,
this matrix is said to be in \emph{canonical form},
and  the new integral candidates $ \vec{ \rJ}$ are called a \emph{canonical basis}~\cite{Henn:2013pwa}. 
While the generalization of a canonical basis beyond polylogarithms in not yet fully understood,
advances have been made  in extending $\epsilon$-factorized bases to arbitrarily complicated
geometries
~\cite{Primo:2017ipr,Frellesvig:2021hkr,Frellesvig:2023iwr,Pogel:2022vat,Pogel:2022yat,Gorges:2023zgv}.

For our problem, we achieved an $\epsilon$-factorization by leveraging many of these developments.
In particular, for the planar topologies, and for all polylogarithmic sub-sectors 
of the non-planar topology,
we used unitarity cuts and multivariate residue analysis~\cite{Henn:2020lye} 
to select integral candidates with unit leading singularities, see also~\cite{Henn:2013woa,Duhr:2021fhk}.
Starting from the six-propagator non-planar integrals generalizations of these methods 
to genus-one geometries become necessary. 
In fact, it is easy to show that the maximal cut of the irreducible six-propagator
non-planar four-point graph (see right panel in fig.~\ref{fig-non-planar-top})
in Baikov representation~\cite{Baikov:1996iu,Frellesvig:2017aai} can be expressed as
\begin{widetext}
\beq
\label{eq:baikov_cut}
\textrm{MaxCut}_{\mathcal{C}}\left[\rI_{110111100}\right]\sim
\int_{\mathcal{C}} \frac{\rd z_2\wedge \rd z_1 }{z_2\sqrt{(z_1-s-z_2)(z_1-s+4 m^2-z_2)}\sqrt{(t z_1-st+s z_2)^2-4m^2(t z_1^2+s (t-z_2)^2)}}\,.
 \eeq
 \end{widetext}
By further taking the residue at $z_2=0$ in~\eqref{eq:baikov_cut}, 
one is left with an integral over a family of  elliptic curves
\beq
 \label{eq:E4-def}
  \mathcal{E}_4: Y^2=(X-e_1)(X-e_2)(X-e_3)(X-e_4)\,,
 \eeq
 with the four roots given by
  \begin{align}
 \label{eq:roots-def}
e_1= &\rs-4\,,\quad e_2=-\frac{\rs \rt +2 \sqrt{\rs\, \rt (\rs+\rt-4)}}{4-\rt}\,,
\nonumber\\
e_3= &-\frac{\rs \rt -2 \sqrt{\rs \,\rt (\rs+\rt-4)}}{4-\rt}\,,\quad e_4=\rs\,.
 \end{align}
We choose as first  \emph{period} for $\mathcal{E}_4$ the integral
 \begin{align}
  \label{eq:periods-def}
 \Psi_0\left(\rs\,,\rt\right)\equiv 2\int_{e_2}^{e_3} \frac{\rd X}{Y}
 =\frac{4 \EK(\lambda)}{\sqrt{(e_1-e_3)(e_2-e_4)}}\,,
  \end{align}
where   $\EK(\lambda)$ is the complete elliptic integral of the first kind 
 and its argument reads 
  \begin{equation}
  \lambda=\frac{4}{2 +\sqrt{\frac{-\rs (\rs+\rt-4)}{-\rt}}}\,.
  \end{equation}

In order to arrive at an $\epsilon$-factorized form, we first notice that all integrals corresponding to the graph of $\rI_{110111100}$ are 
reduced to six independent master integrals (plus subtopologies). We therefore expect two masters integrals which satisfy a coupled differential equation and
map to the generators of the first de Rham cohomology group $H^1_{\textrm{dR}}(\mathcal{E}_4)$, plus four additional ones corresponding to 
independent punctures on the elliptic curve~\cite{Gorges:2023zgv}. 
Candidates for the first two masters can be found for example
starting from the ansatz~\cite{Adams:2018yfj,Adams:2018bsn,Adams:2018kez}
\begin{align}
&\rJ_{47}=\frac{\rI_{110111100}}{\Psi_0} \,, ~~
\rJ_{46} \sim \frac{1}{\eps}\frac{\Psi_0^2}{2\pi i\, W_{\rt}}\partial_{\rt}\rJ_{47}+\dots\,,
 \end{align}
 where $W_{\rt}= \frac{1}{2 \pi i} \frac{1}{ \rt^2 (\rs+\rt)(\rs+\rt-4)}$
is the Wronskian of the second-order Picard-Fuchs equation associated to the elliptic curve.
The explicit expression of $\rJ_{46}$ is immaterial for this discussion, and is given in the supplemental material.
The remaining four candidates can be identified by analysing their integrand representation
and the structure of the resulting differential equations.
As a last step, in order to obtain a fully $\epsilon$-factorized form,
one needs to integrate out some inhomogenous entries in the differential equation matrix, 
which leads to the appearance of additional
transcendental integrals.
In this way, the final $\epsilon$-factorized system~\eqref{eq:eps-form}, is expressed in terms of
$87$ distinct one-forms $\omega_i$.
It is easy to
verify that the integrability condition $\mathrm{d} \mathrm{A}=\mathrm{A}\wedge\mathrm{A}$ 
is satisfied and that  all
${\omega_i }$ are the closed $\rd \omega_i=0$.

The individual differential forms can be  
simplified by exploiting the underlying geometry
of the family of elliptic curves in~(\ref{eq:E4-def}).
As an example, consider  the following two functions
{\small
\begin{align}
\label{eq:T12}
\rT_1(\rs,\rt)=&\int
 \rd \rs\,
 \bigg[\frac{-\rt}{\rs}(4\rs^2+4\rs(\rt-4)+\rt(\rt-4))\Psi_0
 \nonumber\\
-&8 \rt\frac{(\rs+\rt-4)(\rs+\rt)}{ (t+2\rs-4)}\partial_\rs \Psi_0\bigg]
\nonumber\\
 +& \rd \rt
  \bigg[\frac{-\rt}{4-\rt} \frac{-48  +4 \rs+2\rs^2+12\rt+\rs \rt}{ \rt+\rs-4}\Psi_0 \bigg]\,,
 \nonumber\\
 \rT_2(\rs,\rt)=&\sqrt{4-\rt}\sqrt{-\rt}\int
 \rd \rs\,
  \bigg[
  \frac{\rt}{\rs} \frac{4+2 \rs-\rs^2- \rt-\rs t}{2(\rs+\rt-4)}\Psi_0
  \nonumber\\
   -&\frac{1}{2}\rt (1+\rs) \partial_\rs \Psi_0
  \bigg]
  + \sqrt{-\rt}\sqrt{4-\rt} \,
 \rd \rt\bigg[\Psi_0
  \nonumber\\
   \times& \left. \frac{\rs-4}{2(\rs+\rt-4)}
  +\frac{(\rs-4)\rs(1+\rs)}{2(-4+2\rs+\rt)}\partial_\rs \Psi_0
  \right)
    \bigg]\,,
\end{align} }
\hspace{-0.23cm} which are among the objects required
to express the matrix $A$ in~\eqref{eq:eps-form}. Again, formulas are given
assuming $y>4$ and $z<0$ for definiteness.
While the details of the construction are immaterial for
this paper and are discussed elsewhere~\cite{Duhr:2023rki,Bonn:2023geo}, 
it suffices to say that one can parameterize the kinematical variables by
\beq
\label{eq:to-uni-curve}
\rs=2\frac{(1-x)(1+t_4)}{t_4-x}\,,\quad \rt=4 \frac{t_4 (1-x^2)}{x^2-t_4^2}\,,
\eeq 
where $\{[x:Y:1],t_4\}$ are  the \emph{canonical coordinates} 
on the moduli space of elliptic curves given by the variety $Y^2 = (x^2-1)(x^2-t_4)$.  
In these coordinates, 
the period in~(\ref{eq:periods-def}) becomes
\beq \label{eq:psi0HM}
   \Psi_{0}(x,t_4)=\frac{2(x^2-t_4)}{- Y  } \EK(t_4)\,.
 \eeq
We can identifying $t_4$ with a \emph{Hauptmodul} for the congruence subgroup $\Gamma_1(4) \subset \mathrm{SL}_2(\mathbb{Z})$~\cite{maier2008rationally}.
Strikingly, it turns out that by changing variables to the canonical coordinates, one can easily see that the two transcendental integrals in~(\ref{eq:T12}) are just combinations of
simpler functions
{\small \begin{align}
\rT_1(x,t_4)=&8t_4\frac{ \EK(t_4)}{\pi} \bigg[
(1-t_4) \mathcal{F}(x,t_4)
  -  \frac{x^2-1}{ (1+t_4) Y}
 \bigg ]\,,
  \\\nonumber
  \rT_2(x,t_4)=&\frac{1}{\pi} \sqrt{\frac{t_4}{1+t_4}} \frac{t_4 (3-2x)-3x+2}{t_4-x}\EK(t_4)-\frac{f(t_4)}{2 \pi}\,,
\end{align}}
where  $f(t_4)$ is given by 
\beq
\label{eq:inte-over-period}
\partial_{t_4} f=&2 \frac{1-t_4}{\sqrt{t_4} (1+t_4)^{3/2}}\EK(t_4)\,,
\eeq
and $\mathcal{F}(x,t_4)$ is the derivative of the \emph{Abel map}:
\begin{align}
 \label{eq:AbelG4}
\mathcal{F}(x,t_4)=&\EK(t_4) \partial_{t_4}{\left[\frac{1}{\EK(t_4)}
\int_{-1}^x
\frac{\rd X}{\sqrt{(X^2-1)(X^2-t_4)}}
\right]}\,.
\end{align}
Other differential forms in the alphabet can also be substantially simplified
and all double integrals over the periods
can be rewritten in terms of rational functions of $\rT_1$ and $\rT_2$. 
One can then show that
all differential forms are given by combinations of the five algebraic functions
  $\{
 \sqrt{x^2-1} \,,\,\sqrt{x^2-t_4}\,, \,\sqrt{1+t_4}\,,\,\sqrt{t_4}\,,\,\sqrt{1-t_4}
  \}$,
and the three transcendental functions
 $ \{
 \EK(t_4)\,, \,f(t_4)\,, \,\mathcal{F}(x,t_4)
  \}$.
We want to stress that the choice of canonical coordinates in~(\ref{eq:to-uni-curve}) 
is not merely 
an academic curiosity, and the final, simplified form is essential 
to efficiently implement the numerical evaluation of the iterated integrals described below.
To explicitly solve the integrals, 
we first expand~(\ref{eq:g_path_sol}) in $\eps$. At each order, the solution of the differential equation is expressed by Chen iterated integrals~\cite{chen1977iterated} and we fix all boundary conditions
imposing regularities at different phase-space points. In this way we obtain fully analytic results for the non-planar master integrals in terms of Chen iterated integrals.

Currently, there are no public numerical routines to evaluate the special functions that appear in the non-planar sector. We therefore obtain generalized series expanions for all master integrals.
More precisely, we start from the differential equations in $\epsilon$-factorized form in order to algorithmically obtain a small mass expansion for the individual master integrals.
In particular, we obtain a generalized power series 
(including logarithms of the mass), 
whose coefficients can be expressed in terms of harmonic polylogarithms~\cite{Remiddi:1999ew}. 
We obtain results that are valid both for the kinematics relevant
for Bhabha and \moller scattering.
As a cross check, we  compared individual master integrals against a direct numerical evaluation with \texttt{AMFlow}~\cite{Liu:2022chg}, both for Bhabha and \moller scattering kinematics, and found agreement to high precision. 
Our series expansions
allow for fast numerical evaluation, appropriate for phenomenological studies. 
A precise description of the numerical implementations 
can be found in the description of the ancillary files 
along with the arXiv submission of this manuscript.

\section{UV renormalization and  IR factorization }
Using the master integrals calculated above, as well as the planar integrals from~\cite{Henn:2013woa,Duhr:2021fhk}, we can obtain an analytic result for the bare amplitude for both polarized and unpolarized scattering. 
The UV divergences can then be renormalized according to 
\beq
\cA^\textrm{r}(\alpha_e,m,s,t,\eps)=Z_2^2\cA(\alpha^{0}_e,m_b,s,t,\eps)\,,
\eeq
with the relation between bare and physical quantities
\beq
\frac{e^2}{4\pi}= \alpha_e^0= \left(\frac{e^{\gamma_E}}{4\pi}\mu^2\right)^\eps Z_e\alpha_e(\mu)\,,\quad m_b=Z_m m\,.
\eeq
Here $Z_2$ and $Z_m$ are on-shell wave function and mass renormalization constants, 
and $Z_e$ refers to coupling constant renormalization either in the $\overline{\textrm{MS}}$ or on-shell (${\textrm{OS}}$) scheme.
The relevant quantities are collected in the supplemental material.
As expected~\cite{Yennie:1961ad}, after UV renormalization we are left with IR poles which are one-loop-exact,
\beq
\cA^\textrm{OS}(\alpha,m,s,t,\eps)=e^{\frac{\frac{\alpha}{4 \pi} Z_1^{\textrm{IR}}}{\eps}}\cC(\alpha,m,s,t,\eps)\,,
\eeq
where  $\cC$ is the finite remainder function, $\alpha$ is the on-shell electromagnetic coupling,
and $Z_1^{\textrm{IR}}$ is the  anomalous dimension
 which controls the  soft singularities of the amplitude to 
 all-orders through exponentiation~\cite{Becher:2009kw,Ferroglia:2009ep}.
 The exact form of $Z^{\textrm{IR}}$ is immaterial for the present discussion
 and we report it for completeness in the supplemental material.
 
We performed several checks on our results. First of all, we verified that our two-loop amplitudes
have the correct UV and IR behavior, as illustrated above.
In addition, we compared both the bare and 
the finite remainders of our one-loop amplitudes against 
\texttt{OpenLoops}~\cite{Cascioli:2011va,Buccioni:2019sur}
and found perfect agreement. 
We stress here that the unpolarized finite remainders in Conventional Dimensional Regularization equal those in the tHV scheme, while the bare and UV-renormalized amplitudes in general differ. The equality of the finite remainders provides another check of our calculation.


\section{Discussion and conclusions}
\label{sec:conclusions}

Our results for the two-loop amplitudes for Bhabha and \moller scattering are given as generalized series expansion in $x = m/E_{\text{CM}}$. They are provided as computer-readable files in the ancillary material of the arXiv submission for both the polarized and unpolarized scattering amplitudes.
We provide sufficiently high orders to obtain reliable predictions for the low-energy experiments mentioned in the introduction, where we expect the mass effects to be the largest. In the following we discuss some of the phenomenological implications of our results.
We focus here on unpolarized \moller scattering, but all conclusions equally apply to Bhabha scattering.

\begin{figure}[ht!]
  \centering
  \includegraphics[width=0.45\textwidth]{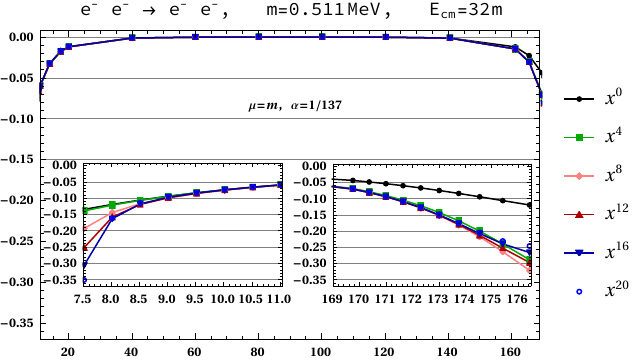}
  \caption{Convergence of the mass expansion.
  Plotted are the 2-loop finite remainders $\cC^{\dagger (2)}\cC^{(0)}$ 
  as functions of scattering angle in degrees, at various truncation orders.
  }
  \label{fig:NNLO-plot}
\end{figure}

Let us start by assessing the accuracy of the small-mass expansion. We begin by noticing that we expect the expansion to become unreliable in the extreme forward or backward regions,
where the coefficients of the series 
in $x$ develop large logarithms in $(-t)/s$ which can 
invalidate the convergence of the expansion.\footnote{This can be interpreted as a manifestation of the lack of commutativity of the small mass limit with the forward limit.}
To quantify the region of convergence, we compare the exact results for the one-loop amplitude $A^{1l}_{\text{exact}}$ 
with the corresponding expansion $A^{1l}_{20}$ to $\mathcal{O}(x^{20})$
and study the ratio $\delta^{1l}_{\text{exact},20} = (A^{1l}_{\text{exact}} - A^{1l}_{20})/A^{1l}_{\text{exact}}$\,.
Depending on the scattering energy $E_{\text{CM}}=\sqrt{s}$, we find that 
$\delta^{1l}_{\text{exact},20} \leq 1\%$
for different ranges of the scattering angle $\theta$:
\begin{alignat}{3} \label{eq:rangestheta}
    &E_{\text{CM}} = 150 m \quad 
    && \to && \quad  2^{\circ}<\theta<179^{\circ}\,, \nonumber \\
    &E_{\text{CM}} = 32 m  \quad 
    && \to && \quad  9^{\circ}<\theta<174^{\circ}\,, \\
    &E_{\text{CM}}  = 5 m  \quad 
    && \to && \quad 70^{\circ}<\theta<130^{\circ}\,, \nonumber
\end{alignat}
where the energy values are chosen to match those probed at present and future experiments.
This shows that at very low energies the expansions
must be interpreted with care outside of the central region.
To extend this to the two-loop amplitudes, 
we repeat the same analysis at one and two loops, 
 comparing this time the series expanded
to order $20$ with the one expanded to order $18$. We find that 
the same applies: for $L=1,2$ $(A^{Ll}_{20} - A^{Ll}_{18})/A^{Ll}_{20} \leq 1\%$ for the same values of $\theta$
as in~\eqref{eq:rangestheta}.
In fig.~\ref{fig:NNLO-plot} we display the various orders of the series
for the two-loop amplitude, for different values of the scattering
at the intermediate energy of $E_{\text{CM}}= 32 m$. We highlight the lack of convergence for $\theta$ not in the range $[9^\circ,174^\circ]$ in the two sub plots.

\begin{figure}[ht!]
  \centering
  \includegraphics[width=0.45\textwidth]{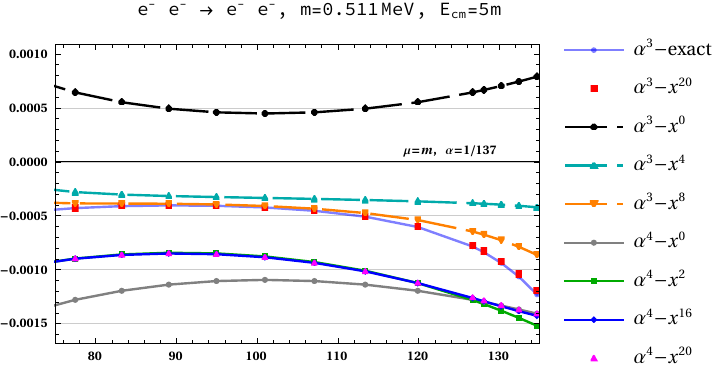}
  \caption{Mass effects at low energies.
  Plotted are the 1-loop $\cC^{\dagger(1)}\cC^{(0)}$ and 2-loop $\cC^{\dagger (2)}\cC^{(0)}$ finite remainders as
  functions of scattering angle in degrees. The 2-loop amplitudes are rescaled by a factor of $25$.}
  \label{fig:mass-effects}
\end{figure}

After having assessed the validity of our small-mass expansions, let us comment on the phenomenological relevance of the mass effects. We only discuss here the mass effects in the purely virtual corrections. So far two-loop mass effects had only been included to leading-power, $\mathcal{O}(x^0)$. We expect that the finite-mass effects are more pronounced for small values of $E_{\text{CM}}$.  In fig.~\ref{fig:NNLO-plot} we see that, for $E_{\text{CM}}=32m$, the two-loop leading-power approximation does not capture the full extend of the mass effects for $\theta\gtrsim 150^\circ$ (for small angles, we are outside the region of~\eqref{eq:rangestheta}). We therefore expect that in that region precise NNLO results can only be obtained by including the subleading terms we have computed. The effect is even more pronounced for $E_{\text{CM}}=5m$: in fig.~\ref{fig:mass-effects} we show that, even in the range of intermediate angles in~\eqref{eq:rangestheta}, the leading-power approximation does not provide a reliable prediction of the finite-mass effects. At the same time, we observe a very nice convergence of the mass expansion, corroborating that we can provide reliable and precise predictions for the two-loop corrections even at such low energies. A full discussion of the size of the NNLO QED corrections will be presented elsewhere.

To conclude, in this letter we have addressed the calculation of the two-loop QED corrections to the scattering 
of four identical massive leptons, retaining full dependence on the lepton mass. This constitutes the last outstanding ingredient
necessary to perform NNLO QED phenomenological studies for standard processes 
as Bhabha and \moller scattering.
In addition to the phenomenological interest behind these calculations, the scattering amplitudes
computed in this paper are an important example of physical processes that receive a non-trivial
contribution from Feynman integrals of elliptic type.
We presented a strategy
to compute these amplitudes analytically through the differential equations method
 and provided a robust numerical implementation. 
 We demonstrated that for low values of $E_{\text{CM}}$, the mass effect can be sizeable and is not captured by the leading-power approximation. We therefore expect that our results will play an important role in making precise predictions for lepton collider experiments possible.

\begin{acknowledgements}
\emph{Acknowledgements:}
We thank Federico Buccioni for providing numerical results for the one-loop amplitudes with \texttt{OpenLoops}
and Christoph Nega for useful comments on the manuscript. 
We are also indebted to Vladimir Smirnov for collaboration in the initial stages of this project.
This work was supported in part by the Excellence
Cluster ORIGINS funded by the Deutsche Forschungsgemeinschaft (DFG, German Research Foundation) under
Germany’s Excellence Strategy – EXC-2094-390783311 
and by the European Research
Council (ERC) under the European Union’s research and
innovation programme grant agreements 949279 (ERC
Starting Grant HighPHun) and 101043686 (ERC Consolidator Grant  LoCoMotive).
Views and opinions expressed
are however those of the author(s) only and do not necessarily reflect those of the
European Union or the European Research Council. Neither the European Union
nor the granting authority can be held responsible for them.

\end{acknowledgements}

\bibliography{bib/exp,bib/four-lep,bib/main,bib/toappear,bib/tools}

\begin{thebibliography}{98}%
\makeatletter
\providecommand \@ifxundefined [1]{%
 \@ifx{#1\undefined}
}%
\providecommand \@ifnum [1]{%
 \ifnum #1\expandafter \@firstoftwo
 \else \expandafter \@secondoftwo
 \fi
}%
\providecommand \@ifx [1]{%
 \ifx #1\expandafter \@firstoftwo
 \else \expandafter \@secondoftwo
 \fi
}%
\providecommand \natexlab [1]{#1}%
\providecommand \enquote  [1]{``#1''}%
\providecommand \bibnamefont  [1]{#1}%
\providecommand \bibfnamefont [1]{#1}%
\providecommand \citenamefont [1]{#1}%
\providecommand \href@noop [0]{\@secondoftwo}%
\providecommand \href [0]{\begingroup \@sanitize@url \@href}%
\providecommand \@href[1]{\@@startlink{#1}\@@href}%
\providecommand \@@href[1]{\endgroup#1\@@endlink}%
\providecommand \@sanitize@url [0]{\catcode `\\12\catcode `\$12\catcode
  `\&12\catcode `\#12\catcode `\^12\catcode `\_12\catcode `\%12\relax}%
\providecommand \@@startlink[1]{}%
\providecommand \@@endlink[0]{}%
\providecommand \url  [0]{\begingroup\@sanitize@url \@url }%
\providecommand \@url [1]{\endgroup\@href {#1}{\urlprefix }}%
\providecommand \urlprefix  [0]{URL }%
\providecommand \Eprint [0]{\href }%
\providecommand \doibase [0]{http://dx.doi.org/}%
\providecommand \selectlanguage [0]{\@gobble}%
\providecommand \bibinfo  [0]{\@secondoftwo}%
\providecommand \bibfield  [0]{\@secondoftwo}%
\providecommand \translation [1]{[#1]}%
\providecommand \BibitemOpen [0]{}%
\providecommand \bibitemStop [0]{}%
\providecommand \bibitemNoStop [0]{.\EOS\space}%
\providecommand \EOS [0]{\spacefactor3000\relax}%
\providecommand \BibitemShut  [1]{\csname bibitem#1\endcsname}%
\let\auto@bib@innerbib\@empty
\bibitem [{\citenamefont {Aguillard}\ \emph {et~al.}(2023)\citenamefont
  {Aguillard} \emph {et~al.}}]{Muong-2:2023cdq}%
  \BibitemOpen
  \bibfield  {author} {\bibinfo {author} {\bibfnamefont {D.~P.}\ \bibnamefont
  {Aguillard}} \emph {et~al.} (\bibinfo {collaboration} {Muon g-2}),\
  }\href@noop {} {\  (\bibinfo {year} {2023})},\ \Eprint
  {http://arxiv.org/abs/2308.06230} {arXiv:2308.06230 [hep-ex]} \BibitemShut
  {NoStop}%
\bibitem [{\citenamefont {Pohl}\ \emph {et~al.}(2010)\citenamefont {Pohl} \emph
  {et~al.}}]{Pohl:2010zza}%
  \BibitemOpen
  \bibfield  {author} {\bibinfo {author} {\bibfnamefont {R.}~\bibnamefont
  {Pohl}} \emph {et~al.},\ }\href {\doibase 10.1038/nature09250} {\bibfield
  {journal} {\bibinfo  {journal} {Nature}\ }\textbf {\bibinfo {volume} {466}},\
  \bibinfo {pages} {213} (\bibinfo {year} {2010})}\BibitemShut {NoStop}%
\bibitem [{\citenamefont {Antognini}\ \emph {et~al.}(2013)\citenamefont
  {Antognini} \emph {et~al.}}]{Antognini:2013txn}%
  \BibitemOpen
  \bibfield  {author} {\bibinfo {author} {\bibfnamefont {A.}~\bibnamefont
  {Antognini}} \emph {et~al.},\ }\href {\doibase 10.1126/science.1230016}
  {\bibfield  {journal} {\bibinfo  {journal} {Science}\ }\textbf {\bibinfo
  {volume} {339}},\ \bibinfo {pages} {417} (\bibinfo {year}
  {2013})}\BibitemShut {NoStop}%
\bibitem [{\citenamefont {Mohr}\ \emph {et~al.}(2008)\citenamefont {Mohr},
  \citenamefont {Taylor},\ and\ \citenamefont {Newell}}]{Mohr:2008fa}%
  \BibitemOpen
  \bibfield  {author} {\bibinfo {author} {\bibfnamefont {P.~J.}\ \bibnamefont
  {Mohr}}, \bibinfo {author} {\bibfnamefont {B.~N.}\ \bibnamefont {Taylor}}, \
  and\ \bibinfo {author} {\bibfnamefont {D.~B.}\ \bibnamefont {Newell}},\
  }\href {\doibase 10.1103/RevModPhys.80.633} {\bibfield  {journal} {\bibinfo
  {journal} {Rev. Mod. Phys.}\ }\textbf {\bibinfo {volume} {80}},\ \bibinfo
  {pages} {633} (\bibinfo {year} {2008})},\ \Eprint
  {http://arxiv.org/abs/0801.0028} {arXiv:0801.0028 [physics.atom-ph]}
  \BibitemShut {NoStop}%
\bibitem [{\citenamefont {Gasparian}\ \emph {et~al.}(2020)\citenamefont
  {Gasparian} \emph {et~al.}}]{PRad:2020oor}%
  \BibitemOpen
  \bibfield  {author} {\bibinfo {author} {\bibfnamefont {A.}~\bibnamefont
  {Gasparian}} \emph {et~al.} (\bibinfo {collaboration} {PRad}),\ }\href@noop
  {} {\  (\bibinfo {year} {2020})},\ \Eprint {http://arxiv.org/abs/2009.10510}
  {arXiv:2009.10510 [nucl-ex]} \BibitemShut {NoStop}%
\bibitem [{\citenamefont {Banerjee}\ \emph {et~al.}(2020)\citenamefont
  {Banerjee}, \citenamefont {Engel}, \citenamefont {Signer},\ and\
  \citenamefont {Ulrich}}]{Banerjee:2020rww}%
  \BibitemOpen
  \bibfield  {author} {\bibinfo {author} {\bibfnamefont {P.}~\bibnamefont
  {Banerjee}}, \bibinfo {author} {\bibfnamefont {T.}~\bibnamefont {Engel}},
  \bibinfo {author} {\bibfnamefont {A.}~\bibnamefont {Signer}}, \ and\ \bibinfo
  {author} {\bibfnamefont {Y.}~\bibnamefont {Ulrich}},\ }\href {\doibase
  10.21468/SciPostPhys.9.2.027} {\bibfield  {journal} {\bibinfo  {journal}
  {SciPost Phys.}\ }\textbf {\bibinfo {volume} {9}},\ \bibinfo {pages} {027}
  (\bibinfo {year} {2020})},\ \Eprint {http://arxiv.org/abs/2007.01654}
  {arXiv:2007.01654 [hep-ph]} \BibitemShut {NoStop}%
\bibitem [{\citenamefont {Banerjee}\ \emph {et~al.}(2022)\citenamefont
  {Banerjee}, \citenamefont {Engel}, \citenamefont {Schalch}, \citenamefont
  {Signer},\ and\ \citenamefont {Ulrich}}]{Banerjee:2021qvi}%
  \BibitemOpen
  \bibfield  {author} {\bibinfo {author} {\bibfnamefont {P.}~\bibnamefont
  {Banerjee}}, \bibinfo {author} {\bibfnamefont {T.}~\bibnamefont {Engel}},
  \bibinfo {author} {\bibfnamefont {N.}~\bibnamefont {Schalch}}, \bibinfo
  {author} {\bibfnamefont {A.}~\bibnamefont {Signer}}, \ and\ \bibinfo {author}
  {\bibfnamefont {Y.}~\bibnamefont {Ulrich}},\ }\href {\doibase
  10.1103/PhysRevD.105.L031904} {\bibfield  {journal} {\bibinfo  {journal}
  {Phys. Rev. D}\ }\textbf {\bibinfo {volume} {105}},\ \bibinfo {pages}
  {L031904} (\bibinfo {year} {2022})},\ \Eprint
  {http://arxiv.org/abs/2107.12311} {arXiv:2107.12311 [hep-ph]} \BibitemShut
  {NoStop}%
\bibitem [{\citenamefont {Banerjee}\ \emph {et~al.}(2021)\citenamefont
  {Banerjee}, \citenamefont {Engel}, \citenamefont {Schalch}, \citenamefont
  {Signer},\ and\ \citenamefont {Ulrich}}]{Banerjee:2021mty}%
  \BibitemOpen
  \bibfield  {author} {\bibinfo {author} {\bibfnamefont {P.}~\bibnamefont
  {Banerjee}}, \bibinfo {author} {\bibfnamefont {T.}~\bibnamefont {Engel}},
  \bibinfo {author} {\bibfnamefont {N.}~\bibnamefont {Schalch}}, \bibinfo
  {author} {\bibfnamefont {A.}~\bibnamefont {Signer}}, \ and\ \bibinfo {author}
  {\bibfnamefont {Y.}~\bibnamefont {Ulrich}},\ }\href {\doibase
  10.1016/j.physletb.2021.136547} {\bibfield  {journal} {\bibinfo  {journal}
  {Phys. Lett. B}\ }\textbf {\bibinfo {volume} {820}},\ \bibinfo {pages}
  {136547} (\bibinfo {year} {2021})},\ \Eprint
  {http://arxiv.org/abs/2106.07469} {arXiv:2106.07469 [hep-ph]} \BibitemShut
  {NoStop}%
\bibitem [{\citenamefont {Bhabha}(1936)}]{Bhabha:1936zz}%
  \BibitemOpen
  \bibfield  {author} {\bibinfo {author} {\bibfnamefont {H.~J.}\ \bibnamefont
  {Bhabha}},\ }\href {\doibase 10.1098/rspa.1936.0046} {\bibfield  {journal}
  {\bibinfo  {journal} {Proc. Roy. Soc. Lond. A}\ }\textbf {\bibinfo {volume}
  {154}},\ \bibinfo {pages} {195} (\bibinfo {year} {1936})}\BibitemShut
  {NoStop}%
\bibitem [{\citenamefont {Benesch}\ \emph {et~al.}(2014)\citenamefont {Benesch}
  \emph {et~al.}}]{MOLLER:2014iki}%
  \BibitemOpen
  \bibfield  {author} {\bibinfo {author} {\bibfnamefont {J.}~\bibnamefont
  {Benesch}} \emph {et~al.} (\bibinfo {collaboration} {MOLLER}),\ }\href@noop
  {} {\  (\bibinfo {year} {2014})},\ \Eprint {http://arxiv.org/abs/1411.4088}
  {arXiv:1411.4088 [nucl-ex]} \BibitemShut {NoStop}%
\bibitem [{\citenamefont {Epstein}\ \emph {et~al.}(2020)\citenamefont {Epstein}
  \emph {et~al.}}]{Epstein:2019gjd}%
  \BibitemOpen
  \bibfield  {author} {\bibinfo {author} {\bibfnamefont {C.~S.}\ \bibnamefont
  {Epstein}} \emph {et~al.},\ }\href {\doibase 10.1103/PhysRevD.102.012006}
  {\bibfield  {journal} {\bibinfo  {journal} {Phys. Rev. D}\ }\textbf {\bibinfo
  {volume} {102}},\ \bibinfo {pages} {012006} (\bibinfo {year} {2020})},\
  \Eprint {http://arxiv.org/abs/1903.09265} {arXiv:1903.09265 [nucl-ex]}
  \BibitemShut {NoStop}%
\bibitem [{\citenamefont {Bern}\ \emph {et~al.}(2001)\citenamefont {Bern},
  \citenamefont {Dixon},\ and\ \citenamefont {Ghinculov}}]{Bern:2000ie}%
  \BibitemOpen
  \bibfield  {author} {\bibinfo {author} {\bibfnamefont {Z.}~\bibnamefont
  {Bern}}, \bibinfo {author} {\bibfnamefont {L.~J.}\ \bibnamefont {Dixon}}, \
  and\ \bibinfo {author} {\bibfnamefont {A.}~\bibnamefont {Ghinculov}},\ }\href
  {\doibase 10.1103/PhysRevD.63.053007} {\bibfield  {journal} {\bibinfo
  {journal} {Phys. Rev. D}\ }\textbf {\bibinfo {volume} {63}},\ \bibinfo
  {pages} {053007} (\bibinfo {year} {2001})},\ \Eprint
  {http://arxiv.org/abs/hep-ph/0010075} {arXiv:hep-ph/0010075} \BibitemShut
  {NoStop}%
\bibitem [{\citenamefont {Jadach}\ \emph {et~al.}(1996)\citenamefont {Jadach}
  \emph {et~al.}}]{Jadach:1996gu}%
  \BibitemOpen
  \bibfield  {author} {\bibinfo {author} {\bibfnamefont {S.}~\bibnamefont
  {Jadach}} \emph {et~al.},\ }in\ \href {\doibase
  10.5170/CERN-1996-001-V-2.229} {\emph {\bibinfo {booktitle} {{CERN Workshop
  on LEP2 Physics (followed by 2nd meeting, 15-16 Jun 1995 and 3rd meeting 2-3
  Nov 1995)}}}}\ (\bibinfo {year} {1996})\ \Eprint
  {http://arxiv.org/abs/hep-ph/9602393} {arXiv:hep-ph/9602393} \BibitemShut
  {NoStop}%
\bibitem [{\citenamefont {Montagna}\ \emph {et~al.}(1998)\citenamefont
  {Montagna}, \citenamefont {Nicrosini},\ and\ \citenamefont
  {Piccinini}}]{Montagna:1998sp}%
  \BibitemOpen
  \bibfield  {author} {\bibinfo {author} {\bibfnamefont {G.}~\bibnamefont
  {Montagna}}, \bibinfo {author} {\bibfnamefont {O.}~\bibnamefont {Nicrosini}},
  \ and\ \bibinfo {author} {\bibfnamefont {F.}~\bibnamefont {Piccinini}},\
  }\href {\doibase 10.1007/BF02845546} {\bibfield  {journal} {\bibinfo
  {journal} {Riv. Nuovo Cim.}\ }\textbf {\bibinfo {volume} {21N9}},\ \bibinfo
  {pages} {1} (\bibinfo {year} {1998})},\ \Eprint
  {http://arxiv.org/abs/hep-ph/9802302} {arXiv:hep-ph/9802302} \BibitemShut
  {NoStop}%
\bibitem [{\citenamefont {Montagna}\ \emph {et~al.}(1999)\citenamefont
  {Montagna}, \citenamefont {Nicrosini}, \citenamefont {Piccinini},\ and\
  \citenamefont {Passarino}}]{Montagna:1998kp}%
  \BibitemOpen
  \bibfield  {author} {\bibinfo {author} {\bibfnamefont {G.}~\bibnamefont
  {Montagna}}, \bibinfo {author} {\bibfnamefont {O.}~\bibnamefont {Nicrosini}},
  \bibinfo {author} {\bibfnamefont {F.}~\bibnamefont {Piccinini}}, \ and\
  \bibinfo {author} {\bibfnamefont {G.}~\bibnamefont {Passarino}},\ }\href
  {\doibase 10.1016/S0010-4655(98)00080-0} {\bibfield  {journal} {\bibinfo
  {journal} {Comput. Phys. Commun.}\ }\textbf {\bibinfo {volume} {117}},\
  \bibinfo {pages} {278} (\bibinfo {year} {1999})},\ \Eprint
  {http://arxiv.org/abs/hep-ph/9804211} {arXiv:hep-ph/9804211} \BibitemShut
  {NoStop}%
\bibitem [{\citenamefont {Penin}(2005)}]{Penin:2005kf}%
  \BibitemOpen
  \bibfield  {author} {\bibinfo {author} {\bibfnamefont {A.~A.}\ \bibnamefont
  {Penin}},\ }\href {\doibase 10.1103/PhysRevLett.95.010408} {\bibfield
  {journal} {\bibinfo  {journal} {Phys. Rev. Lett.}\ }\textbf {\bibinfo
  {volume} {95}},\ \bibinfo {pages} {010408} (\bibinfo {year} {2005})},\
  \Eprint {http://arxiv.org/abs/hep-ph/0501120} {arXiv:hep-ph/0501120}
  \BibitemShut {NoStop}%
\bibitem [{\citenamefont {Penin}(2006)}]{Penin:2005eh}%
  \BibitemOpen
  \bibfield  {author} {\bibinfo {author} {\bibfnamefont {A.~A.}\ \bibnamefont
  {Penin}},\ }\href {\doibase 10.1016/j.nuclphysb.2005.11.016} {\bibfield
  {journal} {\bibinfo  {journal} {Nucl. Phys. B}\ }\textbf {\bibinfo {volume}
  {734}},\ \bibinfo {pages} {185} (\bibinfo {year} {2006})},\ \Eprint
  {http://arxiv.org/abs/hep-ph/0508127} {arXiv:hep-ph/0508127} \BibitemShut
  {NoStop}%
\bibitem [{\citenamefont {Mitov}\ and\ \citenamefont
  {Moch}(2007)}]{Mitov:2006xs}%
  \BibitemOpen
  \bibfield  {author} {\bibinfo {author} {\bibfnamefont {A.}~\bibnamefont
  {Mitov}}\ and\ \bibinfo {author} {\bibfnamefont {S.}~\bibnamefont {Moch}},\
  }\href {\doibase 10.1088/1126-6708/2007/05/001} {\bibfield  {journal}
  {\bibinfo  {journal} {JHEP}\ }\textbf {\bibinfo {volume} {05}},\ \bibinfo
  {pages} {001} (\bibinfo {year} {2007})},\ \Eprint
  {http://arxiv.org/abs/hep-ph/0612149} {arXiv:hep-ph/0612149} \BibitemShut
  {NoStop}%
\bibitem [{\citenamefont {Becher}\ and\ \citenamefont
  {Melnikov}(2007)}]{Becher:2007cu}%
  \BibitemOpen
  \bibfield  {author} {\bibinfo {author} {\bibfnamefont {T.}~\bibnamefont
  {Becher}}\ and\ \bibinfo {author} {\bibfnamefont {K.}~\bibnamefont
  {Melnikov}},\ }\href {\doibase 10.1088/1126-6708/2007/06/084} {\bibfield
  {journal} {\bibinfo  {journal} {JHEP}\ }\textbf {\bibinfo {volume} {06}},\
  \bibinfo {pages} {084} (\bibinfo {year} {2007})},\ \Eprint
  {http://arxiv.org/abs/0704.3582} {arXiv:0704.3582 [hep-ph]} \BibitemShut
  {NoStop}%
\bibitem [{\citenamefont {Actis}\ \emph {et~al.}(2007)\citenamefont {Actis},
  \citenamefont {Czakon}, \citenamefont {Gluza},\ and\ \citenamefont
  {Riemann}}]{Actis:2007gi}%
  \BibitemOpen
  \bibfield  {author} {\bibinfo {author} {\bibfnamefont {S.}~\bibnamefont
  {Actis}}, \bibinfo {author} {\bibfnamefont {M.}~\bibnamefont {Czakon}},
  \bibinfo {author} {\bibfnamefont {J.}~\bibnamefont {Gluza}}, \ and\ \bibinfo
  {author} {\bibfnamefont {T.}~\bibnamefont {Riemann}},\ }\href {\doibase
  10.1016/j.nuclphysb.2007.06.023} {\bibfield  {journal} {\bibinfo  {journal}
  {Nucl. Phys. B}\ }\textbf {\bibinfo {volume} {786}},\ \bibinfo {pages} {26}
  (\bibinfo {year} {2007})},\ \Eprint {http://arxiv.org/abs/0704.2400}
  {arXiv:0704.2400 [hep-ph]} \BibitemShut {NoStop}%
\bibitem [{\citenamefont {Penin}\ and\ \citenamefont
  {Zerf}(2016)}]{Penin:2016wiw}%
  \BibitemOpen
  \bibfield  {author} {\bibinfo {author} {\bibfnamefont {A.~A.}\ \bibnamefont
  {Penin}}\ and\ \bibinfo {author} {\bibfnamefont {N.}~\bibnamefont {Zerf}},\
  }\href {\doibase 10.1016/j.physletb.2016.07.077} {\bibfield  {journal}
  {\bibinfo  {journal} {Phys. Lett. B}\ }\textbf {\bibinfo {volume} {760}},\
  \bibinfo {pages} {816} (\bibinfo {year} {2016})},\ \bibinfo {note} {[Erratum:
  Phys.Lett.B 771, 637--637 (2017)]},\ \Eprint
  {http://arxiv.org/abs/1606.06344} {arXiv:1606.06344 [hep-ph]} \BibitemShut
  {NoStop}%
\bibitem [{\citenamefont {Bonciani}\ \emph {et~al.}(2004)\citenamefont
  {Bonciani}, \citenamefont {Ferroglia}, \citenamefont {Mastrolia},
  \citenamefont {Remiddi},\ and\ \citenamefont {van~der
  Bij}}]{Bonciani:2004gi}%
  \BibitemOpen
  \bibfield  {author} {\bibinfo {author} {\bibfnamefont {R.}~\bibnamefont
  {Bonciani}}, \bibinfo {author} {\bibfnamefont {A.}~\bibnamefont {Ferroglia}},
  \bibinfo {author} {\bibfnamefont {P.}~\bibnamefont {Mastrolia}}, \bibinfo
  {author} {\bibfnamefont {E.}~\bibnamefont {Remiddi}}, \ and\ \bibinfo
  {author} {\bibfnamefont {J.~J.}\ \bibnamefont {van~der Bij}},\ }\href
  {\doibase 10.1016/j.nuclphysb.2004.09.015} {\bibfield  {journal} {\bibinfo
  {journal} {Nucl. Phys. B}\ }\textbf {\bibinfo {volume} {701}},\ \bibinfo
  {pages} {121} (\bibinfo {year} {2004})},\ \Eprint
  {http://arxiv.org/abs/hep-ph/0405275} {arXiv:hep-ph/0405275} \BibitemShut
  {NoStop}%
\bibitem [{\citenamefont {Kuhn}\ \emph {et~al.}(2001)\citenamefont {Kuhn},
  \citenamefont {Moch}, \citenamefont {Penin},\ and\ \citenamefont
  {Smirnov}}]{Kuhn:2001hz}%
  \BibitemOpen
  \bibfield  {author} {\bibinfo {author} {\bibfnamefont {J.~H.}\ \bibnamefont
  {Kuhn}}, \bibinfo {author} {\bibfnamefont {S.}~\bibnamefont {Moch}}, \bibinfo
  {author} {\bibfnamefont {A.~A.}\ \bibnamefont {Penin}}, \ and\ \bibinfo
  {author} {\bibfnamefont {V.~A.}\ \bibnamefont {Smirnov}},\ }\href {\doibase
  10.1016/S0550-3213(01)00454-0} {\bibfield  {journal} {\bibinfo  {journal}
  {Nucl. Phys. B}\ }\textbf {\bibinfo {volume} {616}},\ \bibinfo {pages} {286}
  (\bibinfo {year} {2001})},\ \bibinfo {note} {[Erratum: Nucl.Phys.B 648,
  455--456 (2003)]},\ \Eprint {http://arxiv.org/abs/hep-ph/0106298}
  {arXiv:hep-ph/0106298} \BibitemShut {NoStop}%
\bibitem [{\citenamefont {Feucht}\ \emph {et~al.}(2004)\citenamefont {Feucht},
  \citenamefont {Kuhn}, \citenamefont {Penin},\ and\ \citenamefont
  {Smirnov}}]{Feucht:2004rp}%
  \BibitemOpen
  \bibfield  {author} {\bibinfo {author} {\bibfnamefont {B.}~\bibnamefont
  {Feucht}}, \bibinfo {author} {\bibfnamefont {J.~H.}\ \bibnamefont {Kuhn}},
  \bibinfo {author} {\bibfnamefont {A.~A.}\ \bibnamefont {Penin}}, \ and\
  \bibinfo {author} {\bibfnamefont {V.~A.}\ \bibnamefont {Smirnov}},\ }\href
  {\doibase 10.1103/PhysRevLett.93.101802} {\bibfield  {journal} {\bibinfo
  {journal} {Phys. Rev. Lett.}\ }\textbf {\bibinfo {volume} {93}},\ \bibinfo
  {pages} {101802} (\bibinfo {year} {2004})},\ \Eprint
  {http://arxiv.org/abs/hep-ph/0404082} {arXiv:hep-ph/0404082} \BibitemShut
  {NoStop}%
\bibitem [{\citenamefont {Jantzen}\ \emph {et~al.}(2005)\citenamefont
  {Jantzen}, \citenamefont {Kuhn}, \citenamefont {Penin},\ and\ \citenamefont
  {Smirnov}}]{Jantzen:2005az}%
  \BibitemOpen
  \bibfield  {author} {\bibinfo {author} {\bibfnamefont {B.}~\bibnamefont
  {Jantzen}}, \bibinfo {author} {\bibfnamefont {J.~H.}\ \bibnamefont {Kuhn}},
  \bibinfo {author} {\bibfnamefont {A.~A.}\ \bibnamefont {Penin}}, \ and\
  \bibinfo {author} {\bibfnamefont {V.~A.}\ \bibnamefont {Smirnov}},\ }\href
  {\doibase 10.1016/j.nuclphysb.2005.10.010} {\bibfield  {journal} {\bibinfo
  {journal} {Nucl. Phys. B}\ }\textbf {\bibinfo {volume} {731}},\ \bibinfo
  {pages} {188} (\bibinfo {year} {2005})},\ \bibinfo {note} {[Erratum:
  Nucl.Phys.B 752, 327--328 (2006)]},\ \Eprint
  {http://arxiv.org/abs/hep-ph/0509157} {arXiv:hep-ph/0509157} \BibitemShut
  {NoStop}%
\bibitem [{\citenamefont {Penin}\ and\ \citenamefont
  {Ryan}(2011)}]{Penin:2011aa}%
  \BibitemOpen
  \bibfield  {author} {\bibinfo {author} {\bibfnamefont {A.~A.}\ \bibnamefont
  {Penin}}\ and\ \bibinfo {author} {\bibfnamefont {G.}~\bibnamefont {Ryan}},\
  }\href {\doibase 10.1007/JHEP11(2011)081} {\bibfield  {journal} {\bibinfo
  {journal} {JHEP}\ }\textbf {\bibinfo {volume} {11}},\ \bibinfo {pages} {081}
  (\bibinfo {year} {2011})},\ \Eprint {http://arxiv.org/abs/1112.2171}
  {arXiv:1112.2171 [hep-ph]} \BibitemShut {NoStop}%
\bibitem [{\citenamefont {Smirnov}(2002)}]{Smirnov:2001cm}%
  \BibitemOpen
  \bibfield  {author} {\bibinfo {author} {\bibfnamefont {V.~A.}\ \bibnamefont
  {Smirnov}},\ }\href {\doibase 10.1016/S0370-2693(01)01382-X} {\bibfield
  {journal} {\bibinfo  {journal} {Phys. Lett. B}\ }\textbf {\bibinfo {volume}
  {524}},\ \bibinfo {pages} {129} (\bibinfo {year} {2002})},\ \Eprint
  {http://arxiv.org/abs/hep-ph/0111160} {arXiv:hep-ph/0111160} \BibitemShut
  {NoStop}%
\bibitem [{\citenamefont {Heinrich}\ and\ \citenamefont
  {Smirnov}(2004)}]{Heinrich:2004iq}%
  \BibitemOpen
  \bibfield  {author} {\bibinfo {author} {\bibfnamefont {G.}~\bibnamefont
  {Heinrich}}\ and\ \bibinfo {author} {\bibfnamefont {V.~A.}\ \bibnamefont
  {Smirnov}},\ }\href {\doibase 10.1016/j.physletb.2004.07.058} {\bibfield
  {journal} {\bibinfo  {journal} {Phys. Lett. B}\ }\textbf {\bibinfo {volume}
  {598}},\ \bibinfo {pages} {55} (\bibinfo {year} {2004})},\ \Eprint
  {http://arxiv.org/abs/hep-ph/0406053} {arXiv:hep-ph/0406053} \BibitemShut
  {NoStop}%
\bibitem [{\citenamefont {Czakon}\ \emph
  {et~al.}(2005{\natexlab{a}})\citenamefont {Czakon}, \citenamefont {Gluza},\
  and\ \citenamefont {Riemann}}]{Czakon:2004wm}%
  \BibitemOpen
  \bibfield  {author} {\bibinfo {author} {\bibfnamefont {M.}~\bibnamefont
  {Czakon}}, \bibinfo {author} {\bibfnamefont {J.}~\bibnamefont {Gluza}}, \
  and\ \bibinfo {author} {\bibfnamefont {T.}~\bibnamefont {Riemann}},\ }\href
  {\doibase 10.1103/PhysRevD.71.073009} {\bibfield  {journal} {\bibinfo
  {journal} {Phys. Rev. D}\ }\textbf {\bibinfo {volume} {71}},\ \bibinfo
  {pages} {073009} (\bibinfo {year} {2005}{\natexlab{a}})},\ \Eprint
  {http://arxiv.org/abs/hep-ph/0412164} {arXiv:hep-ph/0412164} \BibitemShut
  {NoStop}%
\bibitem [{\citenamefont {Czakon}\ \emph
  {et~al.}(2005{\natexlab{b}})\citenamefont {Czakon}, \citenamefont {Gluza},\
  and\ \citenamefont {Riemann}}]{Czakon:2005gi}%
  \BibitemOpen
  \bibfield  {author} {\bibinfo {author} {\bibfnamefont {M.}~\bibnamefont
  {Czakon}}, \bibinfo {author} {\bibfnamefont {J.}~\bibnamefont {Gluza}}, \
  and\ \bibinfo {author} {\bibfnamefont {T.}~\bibnamefont {Riemann}},\
  }\href@noop {} {\bibfield  {journal} {\bibinfo  {journal} {Acta Phys. Polon.
  B}\ }\textbf {\bibinfo {volume} {36}},\ \bibinfo {pages} {3319} (\bibinfo
  {year} {2005}{\natexlab{b}})},\ \Eprint {http://arxiv.org/abs/hep-ph/0511187}
  {arXiv:hep-ph/0511187} \BibitemShut {NoStop}%
\bibitem [{\citenamefont {Czakon}\ \emph
  {et~al.}(2006{\natexlab{a}})\citenamefont {Czakon}, \citenamefont {Gluza},\
  and\ \citenamefont {Riemann}}]{Czakon:2005jd}%
  \BibitemOpen
  \bibfield  {author} {\bibinfo {author} {\bibfnamefont {M.}~\bibnamefont
  {Czakon}}, \bibinfo {author} {\bibfnamefont {J.}~\bibnamefont {Gluza}}, \
  and\ \bibinfo {author} {\bibfnamefont {T.}~\bibnamefont {Riemann}},\ }\href
  {\doibase 10.1016/j.nima.2005.11.148} {\bibfield  {journal} {\bibinfo
  {journal} {Nucl. Instrum. Meth. A}\ }\textbf {\bibinfo {volume} {559}},\
  \bibinfo {pages} {265} (\bibinfo {year} {2006}{\natexlab{a}})},\ \Eprint
  {http://arxiv.org/abs/hep-ph/0508212} {arXiv:hep-ph/0508212} \BibitemShut
  {NoStop}%
\bibitem [{\citenamefont {Czakon}\ \emph
  {et~al.}(2006{\natexlab{b}})\citenamefont {Czakon}, \citenamefont {Gluza},
  \citenamefont {Kajda},\ and\ \citenamefont {Riemann}}]{Czakon:2006hb}%
  \BibitemOpen
  \bibfield  {author} {\bibinfo {author} {\bibfnamefont {M.}~\bibnamefont
  {Czakon}}, \bibinfo {author} {\bibfnamefont {J.}~\bibnamefont {Gluza}},
  \bibinfo {author} {\bibfnamefont {K.}~\bibnamefont {Kajda}}, \ and\ \bibinfo
  {author} {\bibfnamefont {T.}~\bibnamefont {Riemann}},\ }\href {\doibase
  10.1016/j.nuclphysbps.2006.03.003} {\bibfield  {journal} {\bibinfo  {journal}
  {Nucl. Phys. B Proc. Suppl.}\ }\textbf {\bibinfo {volume} {157}},\ \bibinfo
  {pages} {16} (\bibinfo {year} {2006}{\natexlab{b}})},\ \Eprint
  {http://arxiv.org/abs/hep-ph/0602102} {arXiv:hep-ph/0602102} \BibitemShut
  {NoStop}%
\bibitem [{\citenamefont {Czakon}\ \emph
  {et~al.}(2006{\natexlab{c}})\citenamefont {Czakon}, \citenamefont {Gluza},\
  and\ \citenamefont {Riemann}}]{Czakon:2006pa}%
  \BibitemOpen
  \bibfield  {author} {\bibinfo {author} {\bibfnamefont {M.}~\bibnamefont
  {Czakon}}, \bibinfo {author} {\bibfnamefont {J.}~\bibnamefont {Gluza}}, \
  and\ \bibinfo {author} {\bibfnamefont {T.}~\bibnamefont {Riemann}},\ }\href
  {\doibase 10.1016/j.nuclphysb.2006.05.033} {\bibfield  {journal} {\bibinfo
  {journal} {Nucl. Phys. B}\ }\textbf {\bibinfo {volume} {751}},\ \bibinfo
  {pages} {1} (\bibinfo {year} {2006}{\natexlab{c}})},\ \Eprint
  {http://arxiv.org/abs/hep-ph/0604101} {arXiv:hep-ph/0604101} \BibitemShut
  {NoStop}%
\bibitem [{\citenamefont {Duhr}\ \emph {et~al.}(2021)\citenamefont {Duhr},
  \citenamefont {Smirnov},\ and\ \citenamefont {Tancredi}}]{Duhr:2021fhk}%
  \BibitemOpen
  \bibfield  {author} {\bibinfo {author} {\bibfnamefont {C.}~\bibnamefont
  {Duhr}}, \bibinfo {author} {\bibfnamefont {V.~A.}\ \bibnamefont {Smirnov}}, \
  and\ \bibinfo {author} {\bibfnamefont {L.}~\bibnamefont {Tancredi}},\ }\href
  {\doibase 10.1007/JHEP09(2021)120} {\bibfield  {journal} {\bibinfo  {journal}
  {JHEP}\ }\textbf {\bibinfo {volume} {09}},\ \bibinfo {pages} {120} (\bibinfo
  {year} {2021})},\ \Eprint {http://arxiv.org/abs/2108.03828} {arXiv:2108.03828
  [hep-ph]} \BibitemShut {NoStop}%
\bibitem [{\citenamefont {Henn}\ and\ \citenamefont
  {Smirnov}(2013)}]{Henn:2013woa}%
  \BibitemOpen
  \bibfield  {author} {\bibinfo {author} {\bibfnamefont {J.~M.}\ \bibnamefont
  {Henn}}\ and\ \bibinfo {author} {\bibfnamefont {V.~A.}\ \bibnamefont
  {Smirnov}},\ }\href {\doibase 10.1007/JHEP11(2013)041} {\bibfield  {journal}
  {\bibinfo  {journal} {JHEP}\ }\textbf {\bibinfo {volume} {11}},\ \bibinfo
  {pages} {041} (\bibinfo {year} {2013})},\ \Eprint
  {http://arxiv.org/abs/1307.4083} {arXiv:1307.4083 [hep-th]} \BibitemShut
  {NoStop}%
\bibitem [{\citenamefont {Broedel}\ \emph {et~al.}(2015)\citenamefont
  {Broedel}, \citenamefont {Mafra}, \citenamefont {Matthes},\ and\
  \citenamefont {Schlotterer}}]{Broedel:2014vla}%
  \BibitemOpen
  \bibfield  {author} {\bibinfo {author} {\bibfnamefont {J.}~\bibnamefont
  {Broedel}}, \bibinfo {author} {\bibfnamefont {C.~R.}\ \bibnamefont {Mafra}},
  \bibinfo {author} {\bibfnamefont {N.}~\bibnamefont {Matthes}}, \ and\
  \bibinfo {author} {\bibfnamefont {O.}~\bibnamefont {Schlotterer}},\ }\href
  {\doibase 10.1007/JHEP07(2015)112} {\bibfield  {journal} {\bibinfo  {journal}
  {JHEP}\ }\textbf {\bibinfo {volume} {07}},\ \bibinfo {pages} {112} (\bibinfo
  {year} {2015})},\ \Eprint {http://arxiv.org/abs/1412.5535} {arXiv:1412.5535
  [hep-th]} \BibitemShut {NoStop}%
\bibitem [{\citenamefont {Adams}\ \emph {et~al.}(2016)\citenamefont {Adams},
  \citenamefont {Bogner}, \citenamefont {Schweitzer},\ and\ \citenamefont
  {Weinzierl}}]{Adams:2016xah}%
  \BibitemOpen
  \bibfield  {author} {\bibinfo {author} {\bibfnamefont {L.}~\bibnamefont
  {Adams}}, \bibinfo {author} {\bibfnamefont {C.}~\bibnamefont {Bogner}},
  \bibinfo {author} {\bibfnamefont {A.}~\bibnamefont {Schweitzer}}, \ and\
  \bibinfo {author} {\bibfnamefont {S.}~\bibnamefont {Weinzierl}},\ }\href
  {\doibase 10.1063/1.4969060} {\bibfield  {journal} {\bibinfo  {journal} {J.
  Math. Phys.}\ }\textbf {\bibinfo {volume} {57}},\ \bibinfo {pages} {122302}
  (\bibinfo {year} {2016})},\ \Eprint {http://arxiv.org/abs/1607.01571}
  {arXiv:1607.01571 [hep-ph]} \BibitemShut {NoStop}%
\bibitem [{\citenamefont {Adams}\ and\ \citenamefont
  {Weinzierl}(2018{\natexlab{a}})}]{Adams:2017ejb}%
  \BibitemOpen
  \bibfield  {author} {\bibinfo {author} {\bibfnamefont {L.}~\bibnamefont
  {Adams}}\ and\ \bibinfo {author} {\bibfnamefont {S.}~\bibnamefont
  {Weinzierl}},\ }\href {\doibase 10.4310/CNTP.2018.v12.n2.a1} {\bibfield
  {journal} {\bibinfo  {journal} {Commun. Num. Theor. Phys.}\ }\textbf
  {\bibinfo {volume} {12}},\ \bibinfo {pages} {193} (\bibinfo {year}
  {2018}{\natexlab{a}})},\ \Eprint {http://arxiv.org/abs/1704.08895}
  {arXiv:1704.08895 [hep-ph]} \BibitemShut {NoStop}%
\bibitem [{\citenamefont {Remiddi}\ and\ \citenamefont
  {Tancredi}(2017)}]{Remiddi:2017har}%
  \BibitemOpen
  \bibfield  {author} {\bibinfo {author} {\bibfnamefont {E.}~\bibnamefont
  {Remiddi}}\ and\ \bibinfo {author} {\bibfnamefont {L.}~\bibnamefont
  {Tancredi}},\ }\href {\doibase 10.1016/j.nuclphysb.2017.10.007} {\bibfield
  {journal} {\bibinfo  {journal} {Nucl. Phys. B}\ }\textbf {\bibinfo {volume}
  {925}},\ \bibinfo {pages} {212} (\bibinfo {year} {2017})},\ \Eprint
  {http://arxiv.org/abs/1709.03622} {arXiv:1709.03622 [hep-ph]} \BibitemShut
  {NoStop}%
\bibitem [{\citenamefont {Broedel}\ \emph {et~al.}(2019)\citenamefont
  {Broedel}, \citenamefont {Duhr}, \citenamefont {Dulat}, \citenamefont
  {Penante},\ and\ \citenamefont {Tancredi}}]{Broedel:2018qkq}%
  \BibitemOpen
  \bibfield  {author} {\bibinfo {author} {\bibfnamefont {J.}~\bibnamefont
  {Broedel}}, \bibinfo {author} {\bibfnamefont {C.}~\bibnamefont {Duhr}},
  \bibinfo {author} {\bibfnamefont {F.}~\bibnamefont {Dulat}}, \bibinfo
  {author} {\bibfnamefont {B.}~\bibnamefont {Penante}}, \ and\ \bibinfo
  {author} {\bibfnamefont {L.}~\bibnamefont {Tancredi}},\ }\href {\doibase
  10.1007/JHEP01(2019)023} {\bibfield  {journal} {\bibinfo  {journal} {JHEP}\
  }\textbf {\bibinfo {volume} {01}},\ \bibinfo {pages} {023} (\bibinfo {year}
  {2019})},\ \Eprint {http://arxiv.org/abs/1809.10698} {arXiv:1809.10698
  [hep-th]} \BibitemShut {NoStop}%
\bibitem [{\citenamefont {Duhr}\ and\ \citenamefont
  {Tancredi}(2020)}]{Duhr:2019rrs}%
  \BibitemOpen
  \bibfield  {author} {\bibinfo {author} {\bibfnamefont {C.}~\bibnamefont
  {Duhr}}\ and\ \bibinfo {author} {\bibfnamefont {L.}~\bibnamefont
  {Tancredi}},\ }\href {\doibase 10.1007/JHEP02(2020)105} {\bibfield  {journal}
  {\bibinfo  {journal} {JHEP}\ }\textbf {\bibinfo {volume} {02}},\ \bibinfo
  {pages} {105} (\bibinfo {year} {2020})},\ \Eprint
  {http://arxiv.org/abs/1912.00077} {arXiv:1912.00077 [hep-th]} \BibitemShut
  {NoStop}%
\bibitem [{\citenamefont {Walden}\ and\ \citenamefont
  {Weinzierl}(2021)}]{Walden:2020odh}%
  \BibitemOpen
  \bibfield  {author} {\bibinfo {author} {\bibfnamefont {M.}~\bibnamefont
  {Walden}}\ and\ \bibinfo {author} {\bibfnamefont {S.}~\bibnamefont
  {Weinzierl}},\ }\href {\doibase 10.1016/j.cpc.2021.108020} {\bibfield
  {journal} {\bibinfo  {journal} {Comput. Phys. Commun.}\ }\textbf {\bibinfo
  {volume} {265}},\ \bibinfo {pages} {108020} (\bibinfo {year} {2021})},\
  \Eprint {http://arxiv.org/abs/2010.05271} {arXiv:2010.05271 [hep-ph]}
  \BibitemShut {NoStop}%
\bibitem [{\citenamefont {Henn}(2013)}]{Henn:2013pwa}%
  \BibitemOpen
  \bibfield  {author} {\bibinfo {author} {\bibfnamefont {J.~M.}\ \bibnamefont
  {Henn}},\ }\href {\doibase 10.1103/PhysRevLett.110.251601} {\bibfield
  {journal} {\bibinfo  {journal} {Phys. Rev. Lett.}\ }\textbf {\bibinfo
  {volume} {110}},\ \bibinfo {pages} {251601} (\bibinfo {year} {2013})},\
  \Eprint {http://arxiv.org/abs/1304.1806} {arXiv:1304.1806 [hep-th]}
  \BibitemShut {NoStop}%
\bibitem [{\citenamefont {Primo}\ and\ \citenamefont
  {Tancredi}(2017{\natexlab{a}})}]{Primo:2016ebd}%
  \BibitemOpen
  \bibfield  {author} {\bibinfo {author} {\bibfnamefont {A.}~\bibnamefont
  {Primo}}\ and\ \bibinfo {author} {\bibfnamefont {L.}~\bibnamefont
  {Tancredi}},\ }\href {\doibase 10.1016/j.nuclphysb.2016.12.021} {\bibfield
  {journal} {\bibinfo  {journal} {Nucl. Phys. B}\ }\textbf {\bibinfo {volume}
  {916}},\ \bibinfo {pages} {94} (\bibinfo {year} {2017}{\natexlab{a}})},\
  \Eprint {http://arxiv.org/abs/1610.08397} {arXiv:1610.08397 [hep-ph]}
  \BibitemShut {NoStop}%
\bibitem [{\citenamefont {Primo}\ and\ \citenamefont
  {Tancredi}(2017{\natexlab{b}})}]{Primo:2017ipr}%
  \BibitemOpen
  \bibfield  {author} {\bibinfo {author} {\bibfnamefont {A.}~\bibnamefont
  {Primo}}\ and\ \bibinfo {author} {\bibfnamefont {L.}~\bibnamefont
  {Tancredi}},\ }\href {\doibase 10.1016/j.nuclphysb.2017.05.018} {\bibfield
  {journal} {\bibinfo  {journal} {Nucl. Phys. B}\ }\textbf {\bibinfo {volume}
  {921}},\ \bibinfo {pages} {316} (\bibinfo {year} {2017}{\natexlab{b}})},\
  \Eprint {http://arxiv.org/abs/1704.05465} {arXiv:1704.05465 [hep-ph]}
  \BibitemShut {NoStop}%
\bibitem [{\citenamefont {Frellesvig}(2022)}]{Frellesvig:2021hkr}%
  \BibitemOpen
  \bibfield  {author} {\bibinfo {author} {\bibfnamefont {H.}~\bibnamefont
  {Frellesvig}},\ }\href {\doibase 10.1007/JHEP03(2022)079} {\bibfield
  {journal} {\bibinfo  {journal} {JHEP}\ }\textbf {\bibinfo {volume} {03}},\
  \bibinfo {pages} {079} (\bibinfo {year} {2022})},\ \Eprint
  {http://arxiv.org/abs/2110.07968} {arXiv:2110.07968 [hep-th]} \BibitemShut
  {NoStop}%
\bibitem [{\citenamefont {Giroux}\ and\ \citenamefont
  {Pokraka}(2023)}]{Giroux:2022wav}%
  \BibitemOpen
  \bibfield  {author} {\bibinfo {author} {\bibfnamefont {M.}~\bibnamefont
  {Giroux}}\ and\ \bibinfo {author} {\bibfnamefont {A.}~\bibnamefont
  {Pokraka}},\ }\href {\doibase 10.1007/JHEP03(2023)155} {\bibfield  {journal}
  {\bibinfo  {journal} {JHEP}\ }\textbf {\bibinfo {volume} {03}},\ \bibinfo
  {pages} {155} (\bibinfo {year} {2023})},\ \Eprint
  {http://arxiv.org/abs/2210.09898} {arXiv:2210.09898 [hep-th]} \BibitemShut
  {NoStop}%
\bibitem [{\citenamefont {Dlapa}\ \emph {et~al.}(2023)\citenamefont {Dlapa},
  \citenamefont {Henn},\ and\ \citenamefont {Wagner}}]{Dlapa:2022wdu}%
  \BibitemOpen
  \bibfield  {author} {\bibinfo {author} {\bibfnamefont {C.}~\bibnamefont
  {Dlapa}}, \bibinfo {author} {\bibfnamefont {J.~M.}\ \bibnamefont {Henn}}, \
  and\ \bibinfo {author} {\bibfnamefont {F.~J.}\ \bibnamefont {Wagner}},\
  }\href {\doibase 10.1007/JHEP08(2023)120} {\bibfield  {journal} {\bibinfo
  {journal} {JHEP}\ }\textbf {\bibinfo {volume} {08}},\ \bibinfo {pages} {120}
  (\bibinfo {year} {2023})},\ \Eprint {http://arxiv.org/abs/2211.16357}
  {arXiv:2211.16357 [hep-ph]} \BibitemShut {NoStop}%
\bibitem [{\citenamefont {P\"ogel}\ \emph {et~al.}(2023)\citenamefont
  {P\"ogel}, \citenamefont {Wang},\ and\ \citenamefont
  {Weinzierl}}]{Pogel:2022vat}%
  \BibitemOpen
  \bibfield  {author} {\bibinfo {author} {\bibfnamefont {S.}~\bibnamefont
  {P\"ogel}}, \bibinfo {author} {\bibfnamefont {X.}~\bibnamefont {Wang}}, \
  and\ \bibinfo {author} {\bibfnamefont {S.}~\bibnamefont {Weinzierl}},\ }\href
  {\doibase 10.1007/JHEP04(2023)117} {\bibfield  {journal} {\bibinfo  {journal}
  {JHEP}\ }\textbf {\bibinfo {volume} {04}},\ \bibinfo {pages} {117} (\bibinfo
  {year} {2023})},\ \Eprint {http://arxiv.org/abs/2212.08908} {arXiv:2212.08908
  [hep-th]} \BibitemShut {NoStop}%
\bibitem [{\citenamefont {Frellesvig}\ and\ \citenamefont
  {Weinzierl}(2023)}]{Frellesvig:2023iwr}%
  \BibitemOpen
  \bibfield  {author} {\bibinfo {author} {\bibfnamefont {H.}~\bibnamefont
  {Frellesvig}}\ and\ \bibinfo {author} {\bibfnamefont {S.}~\bibnamefont
  {Weinzierl}},\ }\href@noop {} {\  (\bibinfo {year} {2023})},\ \Eprint
  {http://arxiv.org/abs/2301.02264} {arXiv:2301.02264 [hep-th]} \BibitemShut
  {NoStop}%
\bibitem [{\citenamefont {G\"orges}\ \emph {et~al.}(2023)\citenamefont
  {G\"orges}, \citenamefont {Nega}, \citenamefont {Tancredi},\ and\
  \citenamefont {Wagner}}]{Gorges:2023zgv}%
  \BibitemOpen
  \bibfield  {author} {\bibinfo {author} {\bibfnamefont {L.}~\bibnamefont
  {G\"orges}}, \bibinfo {author} {\bibfnamefont {C.}~\bibnamefont {Nega}},
  \bibinfo {author} {\bibfnamefont {L.}~\bibnamefont {Tancredi}}, \ and\
  \bibinfo {author} {\bibfnamefont {F.~J.}\ \bibnamefont {Wagner}},\ }\href
  {\doibase 10.1007/JHEP07(2023)206} {\bibfield  {journal} {\bibinfo  {journal}
  {JHEP}\ }\textbf {\bibinfo {volume} {07}},\ \bibinfo {pages} {206} (\bibinfo
  {year} {2023})},\ \Eprint {http://arxiv.org/abs/2305.14090} {arXiv:2305.14090
  [hep-th]} \BibitemShut {NoStop}%
\bibitem [{\citenamefont {Jiang}\ \emph {et~al.}(2023)\citenamefont {Jiang},
  \citenamefont {Wang}, \citenamefont {Yang},\ and\ \citenamefont
  {Zhao}}]{Jiang:2023jmk}%
  \BibitemOpen
  \bibfield  {author} {\bibinfo {author} {\bibfnamefont {X.}~\bibnamefont
  {Jiang}}, \bibinfo {author} {\bibfnamefont {X.}~\bibnamefont {Wang}},
  \bibinfo {author} {\bibfnamefont {L.~L.}\ \bibnamefont {Yang}}, \ and\
  \bibinfo {author} {\bibfnamefont {J.}~\bibnamefont {Zhao}},\ }\href {\doibase
  10.1007/JHEP09(2023)187} {\bibfield  {journal} {\bibinfo  {journal} {JHEP}\
  }\textbf {\bibinfo {volume} {09}},\ \bibinfo {pages} {187} (\bibinfo {year}
  {2023})},\ \Eprint {http://arxiv.org/abs/2305.13951} {arXiv:2305.13951
  [hep-th]} \BibitemShut {NoStop}%
\bibitem [{\citenamefont {Peraro}\ and\ \citenamefont
  {Tancredi}(2019)}]{Peraro:2019cjj}%
  \BibitemOpen
  \bibfield  {author} {\bibinfo {author} {\bibfnamefont {T.}~\bibnamefont
  {Peraro}}\ and\ \bibinfo {author} {\bibfnamefont {L.}~\bibnamefont
  {Tancredi}},\ }\href {\doibase 10.1007/JHEP07(2019)114} {\bibfield  {journal}
  {\bibinfo  {journal} {JHEP}\ }\textbf {\bibinfo {volume} {07}},\ \bibinfo
  {pages} {114} (\bibinfo {year} {2019})},\ \Eprint
  {http://arxiv.org/abs/1906.03298} {arXiv:1906.03298 [hep-ph]} \BibitemShut
  {NoStop}%
\bibitem [{\citenamefont {Peraro}\ and\ \citenamefont
  {Tancredi}(2021)}]{Peraro:2020sfm}%
  \BibitemOpen
  \bibfield  {author} {\bibinfo {author} {\bibfnamefont {T.}~\bibnamefont
  {Peraro}}\ and\ \bibinfo {author} {\bibfnamefont {L.}~\bibnamefont
  {Tancredi}},\ }\href {\doibase 10.1103/PhysRevD.103.054042} {\bibfield
  {journal} {\bibinfo  {journal} {Phys. Rev. D}\ }\textbf {\bibinfo {volume}
  {103}},\ \bibinfo {pages} {054042} (\bibinfo {year} {2021})},\ \Eprint
  {http://arxiv.org/abs/2012.00820} {arXiv:2012.00820 [hep-ph]} \BibitemShut
  {NoStop}%
\bibitem [{\citenamefont {'t~Hooft}\ and\ \citenamefont
  {Veltman}(1972)}]{tHooft:1972tcz}%
  \BibitemOpen
  \bibfield  {author} {\bibinfo {author} {\bibfnamefont {G.}~\bibnamefont
  {'t~Hooft}}\ and\ \bibinfo {author} {\bibfnamefont {M.~J.~G.}\ \bibnamefont
  {Veltman}},\ }\href {\doibase 10.1016/0550-3213(72)90279-9} {\bibfield
  {journal} {\bibinfo  {journal} {Nucl. Phys. B}\ }\textbf {\bibinfo {volume}
  {44}},\ \bibinfo {pages} {189} (\bibinfo {year} {1972})}\BibitemShut
  {NoStop}%
\bibitem [{\citenamefont {Nogueira}(1993)}]{Nogueira:1991ex}%
  \BibitemOpen
  \bibfield  {author} {\bibinfo {author} {\bibfnamefont {P.}~\bibnamefont
  {Nogueira}},\ }\href {\doibase 10.1006/jcph.1993.1074} {\bibfield  {journal}
  {\bibinfo  {journal} {J. Comput. Phys.}\ }\textbf {\bibinfo {volume} {105}},\
  \bibinfo {pages} {279} (\bibinfo {year} {1993})}\BibitemShut {NoStop}%
\bibitem [{\citenamefont {Vermaseren}(2000)}]{Vermaseren:2000nd}%
  \BibitemOpen
  \bibfield  {author} {\bibinfo {author} {\bibfnamefont {J.~A.~M.}\
  \bibnamefont {Vermaseren}},\ }\href@noop {} {\  (\bibinfo {year} {2000})},\
  \Eprint {http://arxiv.org/abs/math-ph/0010025} {arXiv:math-ph/0010025}
  \BibitemShut {NoStop}%
\bibitem [{\citenamefont {Kuipers}\ \emph {et~al.}(2013)\citenamefont
  {Kuipers}, \citenamefont {Ueda}, \citenamefont {Vermaseren},\ and\
  \citenamefont {Vollinga}}]{Kuipers:2012rf}%
  \BibitemOpen
  \bibfield  {author} {\bibinfo {author} {\bibfnamefont {J.}~\bibnamefont
  {Kuipers}}, \bibinfo {author} {\bibfnamefont {T.}~\bibnamefont {Ueda}},
  \bibinfo {author} {\bibfnamefont {J.~A.~M.}\ \bibnamefont {Vermaseren}}, \
  and\ \bibinfo {author} {\bibfnamefont {J.}~\bibnamefont {Vollinga}},\ }\href
  {\doibase 10.1016/j.cpc.2012.12.028} {\bibfield  {journal} {\bibinfo
  {journal} {Comput. Phys. Commun.}\ }\textbf {\bibinfo {volume} {184}},\
  \bibinfo {pages} {1453} (\bibinfo {year} {2013})},\ \Eprint
  {http://arxiv.org/abs/1203.6543} {arXiv:1203.6543 [cs.SC]} \BibitemShut
  {NoStop}%
\bibitem [{\citenamefont {Kuipers}\ \emph {et~al.}(2015)\citenamefont
  {Kuipers}, \citenamefont {Ueda},\ and\ \citenamefont
  {Vermaseren}}]{Kuipers:2013pba}%
  \BibitemOpen
  \bibfield  {author} {\bibinfo {author} {\bibfnamefont {J.}~\bibnamefont
  {Kuipers}}, \bibinfo {author} {\bibfnamefont {T.}~\bibnamefont {Ueda}}, \
  and\ \bibinfo {author} {\bibfnamefont {J.~A.~M.}\ \bibnamefont
  {Vermaseren}},\ }\href {\doibase 10.1016/j.cpc.2014.08.008} {\bibfield
  {journal} {\bibinfo  {journal} {Comput. Phys. Commun.}\ }\textbf {\bibinfo
  {volume} {189}},\ \bibinfo {pages} {1} (\bibinfo {year} {2015})},\ \Eprint
  {http://arxiv.org/abs/1310.7007} {arXiv:1310.7007 [cs.SC]} \BibitemShut
  {NoStop}%
\bibitem [{\citenamefont {Ruijl}\ \emph {et~al.}(2017)\citenamefont {Ruijl},
  \citenamefont {Ueda},\ and\ \citenamefont {Vermaseren}}]{Ruijl:2017dtg}%
  \BibitemOpen
  \bibfield  {author} {\bibinfo {author} {\bibfnamefont {B.}~\bibnamefont
  {Ruijl}}, \bibinfo {author} {\bibfnamefont {T.}~\bibnamefont {Ueda}}, \ and\
  \bibinfo {author} {\bibfnamefont {J.}~\bibnamefont {Vermaseren}},\
  }\href@noop {} {\  (\bibinfo {year} {2017})},\ \Eprint
  {http://arxiv.org/abs/1707.06453} {arXiv:1707.06453 [hep-ph]} \BibitemShut
  {NoStop}%
\bibitem [{\citenamefont {von Manteuffel}\ and\ \citenamefont
  {Studerus}(2012)}]{vonManteuffel:2012np}%
  \BibitemOpen
  \bibfield  {author} {\bibinfo {author} {\bibfnamefont {A.}~\bibnamefont {von
  Manteuffel}}\ and\ \bibinfo {author} {\bibfnamefont {C.}~\bibnamefont
  {Studerus}},\ }\href@noop {} {\  (\bibinfo {year} {2012})},\ \Eprint
  {http://arxiv.org/abs/1201.4330} {arXiv:1201.4330 [hep-ph]} \BibitemShut
  {NoStop}%
\bibitem [{\citenamefont {Maierh\"ofer}\ \emph {et~al.}(2018)\citenamefont
  {Maierh\"ofer}, \citenamefont {Usovitsch},\ and\ \citenamefont
  {Uwer}}]{Maierhofer:2017gsa}%
  \BibitemOpen
  \bibfield  {author} {\bibinfo {author} {\bibfnamefont {P.}~\bibnamefont
  {Maierh\"ofer}}, \bibinfo {author} {\bibfnamefont {J.}~\bibnamefont
  {Usovitsch}}, \ and\ \bibinfo {author} {\bibfnamefont {P.}~\bibnamefont
  {Uwer}},\ }\href {\doibase 10.1016/j.cpc.2018.04.012} {\bibfield  {journal}
  {\bibinfo  {journal} {Comput. Phys. Commun.}\ }\textbf {\bibinfo {volume}
  {230}},\ \bibinfo {pages} {99} (\bibinfo {year} {2018})},\ \Eprint
  {http://arxiv.org/abs/1705.05610} {arXiv:1705.05610 [hep-ph]} \BibitemShut
  {NoStop}%
\bibitem [{\citenamefont {Maierh\"ofer}\ and\ \citenamefont
  {Usovitsch}(2018)}]{Maierhofer:2018gpa}%
  \BibitemOpen
  \bibfield  {author} {\bibinfo {author} {\bibfnamefont {P.}~\bibnamefont
  {Maierh\"ofer}}\ and\ \bibinfo {author} {\bibfnamefont {J.}~\bibnamefont
  {Usovitsch}},\ }\href@noop {} {\  (\bibinfo {year} {2018})},\ \Eprint
  {http://arxiv.org/abs/1812.01491} {arXiv:1812.01491 [hep-ph]} \BibitemShut
  {NoStop}%
\bibitem [{\citenamefont {Klappert}\ \emph {et~al.}(2021)\citenamefont
  {Klappert}, \citenamefont {Lange}, \citenamefont {Maierh\"ofer},\ and\
  \citenamefont {Usovitsch}}]{Klappert:2020nbg}%
  \BibitemOpen
  \bibfield  {author} {\bibinfo {author} {\bibfnamefont {J.}~\bibnamefont
  {Klappert}}, \bibinfo {author} {\bibfnamefont {F.}~\bibnamefont {Lange}},
  \bibinfo {author} {\bibfnamefont {P.}~\bibnamefont {Maierh\"ofer}}, \ and\
  \bibinfo {author} {\bibfnamefont {J.}~\bibnamefont {Usovitsch}},\ }\href
  {\doibase 10.1016/j.cpc.2021.108024} {\bibfield  {journal} {\bibinfo
  {journal} {Comput. Phys. Commun.}\ }\textbf {\bibinfo {volume} {266}},\
  \bibinfo {pages} {108024} (\bibinfo {year} {2021})},\ \Eprint
  {http://arxiv.org/abs/2008.06494} {arXiv:2008.06494 [hep-ph]} \BibitemShut
  {NoStop}%
\bibitem [{\citenamefont {Tkachov}(1981)}]{Tkachov:1981wb}%
  \BibitemOpen
  \bibfield  {author} {\bibinfo {author} {\bibfnamefont {F.~V.}\ \bibnamefont
  {Tkachov}},\ }\href {\doibase 10.1016/0370-2693(81)90288-4} {\bibfield
  {journal} {\bibinfo  {journal} {Phys. Lett. B}\ }\textbf {\bibinfo {volume}
  {100}},\ \bibinfo {pages} {65} (\bibinfo {year} {1981})}\BibitemShut
  {NoStop}%
\bibitem [{\citenamefont {Chetyrkin}\ and\ \citenamefont
  {Tkachov}(1981)}]{Chetyrkin:1981qh}%
  \BibitemOpen
  \bibfield  {author} {\bibinfo {author} {\bibfnamefont {K.~G.}\ \bibnamefont
  {Chetyrkin}}\ and\ \bibinfo {author} {\bibfnamefont {F.~V.}\ \bibnamefont
  {Tkachov}},\ }\href {\doibase 10.1016/0550-3213(81)90199-1} {\bibfield
  {journal} {\bibinfo  {journal} {Nucl. Phys. B}\ }\textbf {\bibinfo {volume}
  {192}},\ \bibinfo {pages} {159} (\bibinfo {year} {1981})}\BibitemShut
  {NoStop}%
\bibitem [{\citenamefont {Laporta}(2000)}]{Laporta:2000dsw}%
  \BibitemOpen
  \bibfield  {author} {\bibinfo {author} {\bibfnamefont {S.}~\bibnamefont
  {Laporta}},\ }\href {\doibase 10.1142/S0217751X00002159} {\bibfield
  {journal} {\bibinfo  {journal} {Int. J. Mod. Phys. A}\ }\textbf {\bibinfo
  {volume} {15}},\ \bibinfo {pages} {5087} (\bibinfo {year} {2000})},\ \Eprint
  {http://arxiv.org/abs/hep-ph/0102033} {arXiv:hep-ph/0102033} \BibitemShut
  {NoStop}%
\bibitem [{\citenamefont {von Manteuffel}\ and\ \citenamefont
  {Schabinger}(2015)}]{vonManteuffel:2014ixa}%
  \BibitemOpen
  \bibfield  {author} {\bibinfo {author} {\bibfnamefont {A.}~\bibnamefont {von
  Manteuffel}}\ and\ \bibinfo {author} {\bibfnamefont {R.~M.}\ \bibnamefont
  {Schabinger}},\ }\href {\doibase 10.1016/j.physletb.2015.03.029} {\bibfield
  {journal} {\bibinfo  {journal} {Phys. Lett. B}\ }\textbf {\bibinfo {volume}
  {744}},\ \bibinfo {pages} {101} (\bibinfo {year} {2015})},\ \Eprint
  {http://arxiv.org/abs/1406.4513} {arXiv:1406.4513 [hep-ph]} \BibitemShut
  {NoStop}%
\bibitem [{\citenamefont {Peraro}(2016)}]{Peraro:2016wsq}%
  \BibitemOpen
  \bibfield  {author} {\bibinfo {author} {\bibfnamefont {T.}~\bibnamefont
  {Peraro}},\ }\href {\doibase 10.1007/JHEP12(2016)030} {\bibfield  {journal}
  {\bibinfo  {journal} {JHEP}\ }\textbf {\bibinfo {volume} {12}},\ \bibinfo
  {pages} {030} (\bibinfo {year} {2016})},\ \Eprint
  {http://arxiv.org/abs/1608.01902} {arXiv:1608.01902 [hep-ph]} \BibitemShut
  {NoStop}%
\bibitem [{\citenamefont {Kotikov}(1991{\natexlab{a}})}]{Kotikov:1990kg}%
  \BibitemOpen
  \bibfield  {author} {\bibinfo {author} {\bibfnamefont {A.~V.}\ \bibnamefont
  {Kotikov}},\ }\href {\doibase 10.1016/0370-2693(91)90413-K} {\bibfield
  {journal} {\bibinfo  {journal} {Phys. Lett. B}\ }\textbf {\bibinfo {volume}
  {254}},\ \bibinfo {pages} {158} (\bibinfo {year}
  {1991}{\natexlab{a}})}\BibitemShut {NoStop}%
\bibitem [{\citenamefont {Kotikov}(1991{\natexlab{b}})}]{Kotikov:1991hm}%
  \BibitemOpen
  \bibfield  {author} {\bibinfo {author} {\bibfnamefont {A.~V.}\ \bibnamefont
  {Kotikov}},\ }\href {\doibase 10.1016/0370-2693(91)90834-D} {\bibfield
  {journal} {\bibinfo  {journal} {Phys. Lett. B}\ }\textbf {\bibinfo {volume}
  {259}},\ \bibinfo {pages} {314} (\bibinfo {year}
  {1991}{\natexlab{b}})}\BibitemShut {NoStop}%
\bibitem [{\citenamefont {Kotikov}(1991{\natexlab{c}})}]{Kotikov:1991pm}%
  \BibitemOpen
  \bibfield  {author} {\bibinfo {author} {\bibfnamefont {A.~V.}\ \bibnamefont
  {Kotikov}},\ }\href {\doibase 10.1016/0370-2693(91)90536-Y} {\bibfield
  {journal} {\bibinfo  {journal} {Phys. Lett. B}\ }\textbf {\bibinfo {volume}
  {267}},\ \bibinfo {pages} {123} (\bibinfo {year} {1991}{\natexlab{c}})},\
  \bibinfo {note} {[Erratum: Phys.Lett.B 295, 409--409 (1992)]}\BibitemShut
  {NoStop}%
\bibitem [{\citenamefont {Remiddi}(1997)}]{Remiddi:1997ny}%
  \BibitemOpen
  \bibfield  {author} {\bibinfo {author} {\bibfnamefont {E.}~\bibnamefont
  {Remiddi}},\ }\href {\doibase 10.1007/BF03185566} {\bibfield  {journal}
  {\bibinfo  {journal} {Nuovo Cim. A}\ }\textbf {\bibinfo {volume} {110}},\
  \bibinfo {pages} {1435} (\bibinfo {year} {1997})},\ \Eprint
  {http://arxiv.org/abs/hep-th/9711188} {arXiv:hep-th/9711188} \BibitemShut
  {NoStop}%
\bibitem [{\citenamefont {Gehrmann}\ and\ \citenamefont
  {Remiddi}(2000)}]{Gehrmann:1999as}%
  \BibitemOpen
  \bibfield  {author} {\bibinfo {author} {\bibfnamefont {T.}~\bibnamefont
  {Gehrmann}}\ and\ \bibinfo {author} {\bibfnamefont {E.}~\bibnamefont
  {Remiddi}},\ }\href {\doibase 10.1016/S0550-3213(00)00223-6} {\bibfield
  {journal} {\bibinfo  {journal} {Nucl. Phys. B}\ }\textbf {\bibinfo {volume}
  {580}},\ \bibinfo {pages} {485} (\bibinfo {year} {2000})},\ \Eprint
  {http://arxiv.org/abs/hep-ph/9912329} {arXiv:hep-ph/9912329} \BibitemShut
  {NoStop}%
\bibitem [{\citenamefont {P\"ogel}\ \emph {et~al.}(2022)\citenamefont
  {P\"ogel}, \citenamefont {Wang},\ and\ \citenamefont
  {Weinzierl}}]{Pogel:2022yat}%
  \BibitemOpen
  \bibfield  {author} {\bibinfo {author} {\bibfnamefont {S.}~\bibnamefont
  {P\"ogel}}, \bibinfo {author} {\bibfnamefont {X.}~\bibnamefont {Wang}}, \
  and\ \bibinfo {author} {\bibfnamefont {S.}~\bibnamefont {Weinzierl}},\ }\href
  {\doibase 10.1007/JHEP09(2022)062} {\bibfield  {journal} {\bibinfo  {journal}
  {JHEP}\ }\textbf {\bibinfo {volume} {09}},\ \bibinfo {pages} {062} (\bibinfo
  {year} {2022})},\ \Eprint {http://arxiv.org/abs/2207.12893} {arXiv:2207.12893
  [hep-th]} \BibitemShut {NoStop}%
\bibitem [{\citenamefont {Henn}\ \emph {et~al.}(2020)\citenamefont {Henn},
  \citenamefont {Mistlberger}, \citenamefont {Smirnov},\ and\ \citenamefont
  {Wasser}}]{Henn:2020lye}%
  \BibitemOpen
  \bibfield  {author} {\bibinfo {author} {\bibfnamefont {J.}~\bibnamefont
  {Henn}}, \bibinfo {author} {\bibfnamefont {B.}~\bibnamefont {Mistlberger}},
  \bibinfo {author} {\bibfnamefont {V.~A.}\ \bibnamefont {Smirnov}}, \ and\
  \bibinfo {author} {\bibfnamefont {P.}~\bibnamefont {Wasser}},\ }\href
  {\doibase 10.1007/JHEP04(2020)167} {\bibfield  {journal} {\bibinfo  {journal}
  {JHEP}\ }\textbf {\bibinfo {volume} {04}},\ \bibinfo {pages} {167} (\bibinfo
  {year} {2020})},\ \Eprint {http://arxiv.org/abs/2002.09492} {arXiv:2002.09492
  [hep-ph]} \BibitemShut {NoStop}%
\bibitem [{\citenamefont {Baikov}(1997)}]{Baikov:1996iu}%
  \BibitemOpen
  \bibfield  {author} {\bibinfo {author} {\bibfnamefont {P.~A.}\ \bibnamefont
  {Baikov}},\ }\href {\doibase 10.1016/S0168-9002(97)00126-5} {\bibfield
  {journal} {\bibinfo  {journal} {Nucl. Instrum. Meth. A}\ }\textbf {\bibinfo
  {volume} {389}},\ \bibinfo {pages} {347} (\bibinfo {year} {1997})},\ \Eprint
  {http://arxiv.org/abs/hep-ph/9611449} {arXiv:hep-ph/9611449} \BibitemShut
  {NoStop}%
\bibitem [{\citenamefont {Frellesvig}\ and\ \citenamefont
  {Papadopoulos}(2017)}]{Frellesvig:2017aai}%
  \BibitemOpen
  \bibfield  {author} {\bibinfo {author} {\bibfnamefont {H.}~\bibnamefont
  {Frellesvig}}\ and\ \bibinfo {author} {\bibfnamefont {C.~G.}\ \bibnamefont
  {Papadopoulos}},\ }\href {\doibase 10.1007/JHEP04(2017)083} {\bibfield
  {journal} {\bibinfo  {journal} {JHEP}\ }\textbf {\bibinfo {volume} {04}},\
  \bibinfo {pages} {083} (\bibinfo {year} {2017})},\ \Eprint
  {http://arxiv.org/abs/1701.07356} {arXiv:1701.07356 [hep-ph]} \BibitemShut
  {NoStop}%
\bibitem [{\citenamefont {Adams}\ and\ \citenamefont
  {Weinzierl}(2018{\natexlab{b}})}]{Adams:2018yfj}%
  \BibitemOpen
  \bibfield  {author} {\bibinfo {author} {\bibfnamefont {L.}~\bibnamefont
  {Adams}}\ and\ \bibinfo {author} {\bibfnamefont {S.}~\bibnamefont
  {Weinzierl}},\ }\href {\doibase 10.1016/j.physletb.2018.04.002} {\bibfield
  {journal} {\bibinfo  {journal} {Phys. Lett. B}\ }\textbf {\bibinfo {volume}
  {781}},\ \bibinfo {pages} {270} (\bibinfo {year} {2018}{\natexlab{b}})},\
  \Eprint {http://arxiv.org/abs/1802.05020} {arXiv:1802.05020 [hep-ph]}
  \BibitemShut {NoStop}%
\bibitem [{\citenamefont {Adams}\ \emph
  {et~al.}(2018{\natexlab{a}})\citenamefont {Adams}, \citenamefont {Chaubey},\
  and\ \citenamefont {Weinzierl}}]{Adams:2018bsn}%
  \BibitemOpen
  \bibfield  {author} {\bibinfo {author} {\bibfnamefont {L.}~\bibnamefont
  {Adams}}, \bibinfo {author} {\bibfnamefont {E.}~\bibnamefont {Chaubey}}, \
  and\ \bibinfo {author} {\bibfnamefont {S.}~\bibnamefont {Weinzierl}},\ }\href
  {\doibase 10.1103/PhysRevLett.121.142001} {\bibfield  {journal} {\bibinfo
  {journal} {Phys. Rev. Lett.}\ }\textbf {\bibinfo {volume} {121}},\ \bibinfo
  {pages} {142001} (\bibinfo {year} {2018}{\natexlab{a}})},\ \Eprint
  {http://arxiv.org/abs/1804.11144} {arXiv:1804.11144 [hep-ph]} \BibitemShut
  {NoStop}%
\bibitem [{\citenamefont {Adams}\ \emph
  {et~al.}(2018{\natexlab{b}})\citenamefont {Adams}, \citenamefont {Chaubey},\
  and\ \citenamefont {Weinzierl}}]{Adams:2018kez}%
  \BibitemOpen
  \bibfield  {author} {\bibinfo {author} {\bibfnamefont {L.}~\bibnamefont
  {Adams}}, \bibinfo {author} {\bibfnamefont {E.}~\bibnamefont {Chaubey}}, \
  and\ \bibinfo {author} {\bibfnamefont {S.}~\bibnamefont {Weinzierl}},\ }\href
  {\doibase 10.1007/JHEP10(2018)206} {\bibfield  {journal} {\bibinfo  {journal}
  {JHEP}\ }\textbf {\bibinfo {volume} {10}},\ \bibinfo {pages} {206} (\bibinfo
  {year} {2018}{\natexlab{b}})},\ \Eprint {http://arxiv.org/abs/1806.04981}
  {arXiv:1806.04981 [hep-ph]} \BibitemShut {NoStop}%
\bibitem [{\citenamefont {Duhr}\ and\ \citenamefont
  {Zhu}(2023)}]{Duhr:2023rki}%
  \BibitemOpen
  \bibfield  {author} {\bibinfo {author} {\bibfnamefont {C.}~\bibnamefont
  {Duhr}}\ and\ \bibinfo {author} {\bibfnamefont {Y.~J.}\ \bibnamefont {Zhu}}\
  }(\bibinfo {year} {2023})\ \Eprint {http://arxiv.org/abs/2310.00485}
  {arXiv:2310.00485 [hep-th]} \BibitemShut {NoStop}%
\bibitem [{\citenamefont {Delto}\ \emph {et~al.}(2023)\citenamefont {Delto},
  \citenamefont {Duhr}, \citenamefont {Tancredi},\ and\ \citenamefont
  {Zhu}}]{Bonn:2023geo}%
  \BibitemOpen
  \bibfield  {author} {\bibinfo {author} {\bibfnamefont {M.}~\bibnamefont
  {Delto}}, \bibinfo {author} {\bibfnamefont {C.}~\bibnamefont {Duhr}},
  \bibinfo {author} {\bibfnamefont {L.}~\bibnamefont {Tancredi}}, \ and\
  \bibinfo {author} {\bibfnamefont {Y.~J.}\ \bibnamefont {Zhu}},\ }\href@noop
  {} {\bibfield  {journal} {\bibinfo  {journal} {to appear}\ } (\bibinfo {year}
  {2023})}\BibitemShut {NoStop}%
\bibitem [{\citenamefont {Maier}(2008)}]{maier2008rationally}%
  \BibitemOpen
  \bibfield  {author} {\bibinfo {author} {\bibfnamefont {R.~S.}\ \bibnamefont
  {Maier}},\ }\href@noop {} {\enquote {\bibinfo {title} {On rationally
  parametrized modular equations},}\ } (\bibinfo {year} {2008}),\ \Eprint
  {http://arxiv.org/abs/math/0611041} {arXiv:math/0611041 [math.NT]}
  \BibitemShut {NoStop}%
\bibitem [{\citenamefont {Chen}(1977)}]{chen1977iterated}%
  \BibitemOpen
  \bibfield  {author} {\bibinfo {author} {\bibfnamefont {K.-T.}\ \bibnamefont
  {Chen}},\ }\href@noop {} {\bibfield  {journal} {\bibinfo  {journal} {Bulletin
  of the American Mathematical Society}\ }\textbf {\bibinfo {volume} {83}},\
  \bibinfo {pages} {831} (\bibinfo {year} {1977})}\BibitemShut {NoStop}%
\bibitem [{\citenamefont {Remiddi}\ and\ \citenamefont
  {Vermaseren}(2000)}]{Remiddi:1999ew}%
  \BibitemOpen
  \bibfield  {author} {\bibinfo {author} {\bibfnamefont {E.}~\bibnamefont
  {Remiddi}}\ and\ \bibinfo {author} {\bibfnamefont {J.~A.~M.}\ \bibnamefont
  {Vermaseren}},\ }\href {\doibase 10.1142/S0217751X00000367} {\bibfield
  {journal} {\bibinfo  {journal} {Int. J. Mod. Phys. A}\ }\textbf {\bibinfo
  {volume} {15}},\ \bibinfo {pages} {725} (\bibinfo {year} {2000})},\ \Eprint
  {http://arxiv.org/abs/hep-ph/9905237} {arXiv:hep-ph/9905237} \BibitemShut
  {NoStop}%
\bibitem [{\citenamefont {Liu}\ and\ \citenamefont {Ma}(2023)}]{Liu:2022chg}%
  \BibitemOpen
  \bibfield  {author} {\bibinfo {author} {\bibfnamefont {X.}~\bibnamefont
  {Liu}}\ and\ \bibinfo {author} {\bibfnamefont {Y.-Q.}\ \bibnamefont {Ma}},\
  }\href {\doibase 10.1016/j.cpc.2022.108565} {\bibfield  {journal} {\bibinfo
  {journal} {Comput. Phys. Commun.}\ }\textbf {\bibinfo {volume} {283}},\
  \bibinfo {pages} {108565} (\bibinfo {year} {2023})},\ \Eprint
  {http://arxiv.org/abs/2201.11669} {arXiv:2201.11669 [hep-ph]} \BibitemShut
  {NoStop}%
\bibitem [{\citenamefont {Yennie}\ \emph {et~al.}(1961)\citenamefont {Yennie},
  \citenamefont {Frautschi},\ and\ \citenamefont {Suura}}]{Yennie:1961ad}%
  \BibitemOpen
  \bibfield  {author} {\bibinfo {author} {\bibfnamefont {D.~R.}\ \bibnamefont
  {Yennie}}, \bibinfo {author} {\bibfnamefont {S.~C.}\ \bibnamefont
  {Frautschi}}, \ and\ \bibinfo {author} {\bibfnamefont {H.}~\bibnamefont
  {Suura}},\ }\href {\doibase 10.1016/0003-4916(61)90151-8} {\bibfield
  {journal} {\bibinfo  {journal} {Annals Phys.}\ }\textbf {\bibinfo {volume}
  {13}},\ \bibinfo {pages} {379} (\bibinfo {year} {1961})}\BibitemShut
  {NoStop}%
\bibitem [{\citenamefont {Becher}\ and\ \citenamefont
  {Neubert}(2009)}]{Becher:2009kw}%
  \BibitemOpen
  \bibfield  {author} {\bibinfo {author} {\bibfnamefont {T.}~\bibnamefont
  {Becher}}\ and\ \bibinfo {author} {\bibfnamefont {M.}~\bibnamefont
  {Neubert}},\ }\href {\doibase 10.1103/PhysRevD.79.125004} {\bibfield
  {journal} {\bibinfo  {journal} {Phys. Rev. D}\ }\textbf {\bibinfo {volume}
  {79}},\ \bibinfo {pages} {125004} (\bibinfo {year} {2009})},\ \bibinfo {note}
  {[Erratum: Phys.Rev.D 80, 109901 (2009)]},\ \Eprint
  {http://arxiv.org/abs/0904.1021} {arXiv:0904.1021 [hep-ph]} \BibitemShut
  {NoStop}%
\bibitem [{\citenamefont {Ferroglia}\ \emph {et~al.}(2009)\citenamefont
  {Ferroglia}, \citenamefont {Neubert}, \citenamefont {Pecjak},\ and\
  \citenamefont {Yang}}]{Ferroglia:2009ep}%
  \BibitemOpen
  \bibfield  {author} {\bibinfo {author} {\bibfnamefont {A.}~\bibnamefont
  {Ferroglia}}, \bibinfo {author} {\bibfnamefont {M.}~\bibnamefont {Neubert}},
  \bibinfo {author} {\bibfnamefont {B.~D.}\ \bibnamefont {Pecjak}}, \ and\
  \bibinfo {author} {\bibfnamefont {L.~L.}\ \bibnamefont {Yang}},\ }\href
  {\doibase 10.1103/PhysRevLett.103.201601} {\bibfield  {journal} {\bibinfo
  {journal} {Phys. Rev. Lett.}\ }\textbf {\bibinfo {volume} {103}},\ \bibinfo
  {pages} {201601} (\bibinfo {year} {2009})},\ \Eprint
  {http://arxiv.org/abs/0907.4791} {arXiv:0907.4791 [hep-ph]} \BibitemShut
  {NoStop}%
\bibitem [{\citenamefont {Cascioli}\ \emph {et~al.}(2012)\citenamefont
  {Cascioli}, \citenamefont {Maierhofer},\ and\ \citenamefont
  {Pozzorini}}]{Cascioli:2011va}%
  \BibitemOpen
  \bibfield  {author} {\bibinfo {author} {\bibfnamefont {F.}~\bibnamefont
  {Cascioli}}, \bibinfo {author} {\bibfnamefont {P.}~\bibnamefont
  {Maierhofer}}, \ and\ \bibinfo {author} {\bibfnamefont {S.}~\bibnamefont
  {Pozzorini}},\ }\href {\doibase 10.1103/PhysRevLett.108.111601} {\bibfield
  {journal} {\bibinfo  {journal} {Phys. Rev. Lett.}\ }\textbf {\bibinfo
  {volume} {108}},\ \bibinfo {pages} {111601} (\bibinfo {year} {2012})},\
  \Eprint {http://arxiv.org/abs/1111.5206} {arXiv:1111.5206 [hep-ph]}
  \BibitemShut {NoStop}%
\bibitem [{\citenamefont {Buccioni}\ \emph {et~al.}(2019)\citenamefont
  {Buccioni}, \citenamefont {Lang}, \citenamefont {Lindert}, \citenamefont
  {Maierh\"ofer}, \citenamefont {Pozzorini}, \citenamefont {Zhang},\ and\
  \citenamefont {Zoller}}]{Buccioni:2019sur}%
  \BibitemOpen
  \bibfield  {author} {\bibinfo {author} {\bibfnamefont {F.}~\bibnamefont
  {Buccioni}}, \bibinfo {author} {\bibfnamefont {J.-N.}\ \bibnamefont {Lang}},
  \bibinfo {author} {\bibfnamefont {J.~M.}\ \bibnamefont {Lindert}}, \bibinfo
  {author} {\bibfnamefont {P.}~\bibnamefont {Maierh\"ofer}}, \bibinfo {author}
  {\bibfnamefont {S.}~\bibnamefont {Pozzorini}}, \bibinfo {author}
  {\bibfnamefont {H.}~\bibnamefont {Zhang}}, \ and\ \bibinfo {author}
  {\bibfnamefont {M.~F.}\ \bibnamefont {Zoller}} (\bibinfo {collaboration}
  {OpenLoops 2}),\ }\href {\doibase 10.1140/epjc/s10052-019-7306-2} {\bibfield
  {journal} {\bibinfo  {journal} {Eur. Phys. J. C}\ }\textbf {\bibinfo {volume}
  {79}},\ \bibinfo {pages} {866} (\bibinfo {year} {2019})},\ \Eprint
  {http://arxiv.org/abs/1907.13071} {arXiv:1907.13071 [hep-ph]} \BibitemShut
  {NoStop}%
\bibitem [{\citenamefont {Broadhurst}\ \emph {et~al.}(1991)\citenamefont
  {Broadhurst}, \citenamefont {Gray},\ and\ \citenamefont
  {Schilcher}}]{Broadhurst:1991fy}%
  \BibitemOpen
  \bibfield  {author} {\bibinfo {author} {\bibfnamefont {D.~J.}\ \bibnamefont
  {Broadhurst}}, \bibinfo {author} {\bibfnamefont {N.}~\bibnamefont {Gray}}, \
  and\ \bibinfo {author} {\bibfnamefont {K.}~\bibnamefont {Schilcher}},\ }\href
  {\doibase 10.1007/BF01412333} {\bibfield  {journal} {\bibinfo  {journal} {Z.
  Phys. C}\ }\textbf {\bibinfo {volume} {52}},\ \bibinfo {pages} {111}
  (\bibinfo {year} {1991})}\BibitemShut {NoStop}%
\bibitem [{\citenamefont {Melnikov}\ and\ \citenamefont {van
  Ritbergen}(2000)}]{Melnikov:2000zc}%
  \BibitemOpen
  \bibfield  {author} {\bibinfo {author} {\bibfnamefont {K.}~\bibnamefont
  {Melnikov}}\ and\ \bibinfo {author} {\bibfnamefont {T.}~\bibnamefont {van
  Ritbergen}},\ }\href {\doibase 10.1016/S0550-3213(00)00526-5} {\bibfield
  {journal} {\bibinfo  {journal} {Nucl. Phys. B}\ }\textbf {\bibinfo {volume}
  {591}},\ \bibinfo {pages} {515} (\bibinfo {year} {2000})},\ \Eprint
  {http://arxiv.org/abs/hep-ph/0005131} {arXiv:hep-ph/0005131} \BibitemShut
  {NoStop}%
\bibitem [{\citenamefont {Czakon}\ \emph {et~al.}(2007)\citenamefont {Czakon},
  \citenamefont {Mitov},\ and\ \citenamefont {Moch}}]{Czakon:2007ej}%
  \BibitemOpen
  \bibfield  {author} {\bibinfo {author} {\bibfnamefont {M.}~\bibnamefont
  {Czakon}}, \bibinfo {author} {\bibfnamefont {A.}~\bibnamefont {Mitov}}, \
  and\ \bibinfo {author} {\bibfnamefont {S.}~\bibnamefont {Moch}},\ }\href
  {\doibase 10.1016/j.physletb.2007.06.020} {\bibfield  {journal} {\bibinfo
  {journal} {Phys. Lett. B}\ }\textbf {\bibinfo {volume} {651}},\ \bibinfo
  {pages} {147} (\bibinfo {year} {2007})},\ \Eprint
  {http://arxiv.org/abs/0705.1975} {arXiv:0705.1975 [hep-ph]} \BibitemShut
  {NoStop}%
\bibitem [{\citenamefont {B\"arnreuther}\ \emph {et~al.}(2014)\citenamefont
  {B\"arnreuther}, \citenamefont {Czakon},\ and\ \citenamefont
  {Fiedler}}]{Barnreuther:2013qvf}%
  \BibitemOpen
  \bibfield  {author} {\bibinfo {author} {\bibfnamefont {P.}~\bibnamefont
  {B\"arnreuther}}, \bibinfo {author} {\bibfnamefont {M.}~\bibnamefont
  {Czakon}}, \ and\ \bibinfo {author} {\bibfnamefont {P.}~\bibnamefont
  {Fiedler}},\ }\href {\doibase 10.1007/JHEP02(2014)078} {\bibfield  {journal}
  {\bibinfo  {journal} {JHEP}\ }\textbf {\bibinfo {volume} {02}},\ \bibinfo
  {pages} {078} (\bibinfo {year} {2014})},\ \Eprint
  {http://arxiv.org/abs/1312.6279} {arXiv:1312.6279 [hep-ph]} \BibitemShut
  {NoStop}%
\bibitem [{\citenamefont {Bonciani}\ \emph {et~al.}(2022)\citenamefont
  {Bonciani} \emph {et~al.}}]{Bonciani:2021okt}%
  \BibitemOpen
  \bibfield  {author} {\bibinfo {author} {\bibfnamefont {R.}~\bibnamefont
  {Bonciani}} \emph {et~al.},\ }\href {\doibase 10.1103/PhysRevLett.128.022002}
  {\bibfield  {journal} {\bibinfo  {journal} {Phys. Rev. Lett.}\ }\textbf
  {\bibinfo {volume} {128}},\ \bibinfo {pages} {022002} (\bibinfo {year}
  {2022})},\ \Eprint {http://arxiv.org/abs/2106.13179} {arXiv:2106.13179
  [hep-ph]} \BibitemShut {NoStop}%
\bibitem [{\citenamefont {Grozin}(2005)}]{Grozin:2005yg}%
  \BibitemOpen
  \bibfield  {author} {\bibinfo {author} {\bibfnamefont {A.}~\bibnamefont
  {Grozin}},\ }in\ \href@noop {} {\emph {\bibinfo {booktitle} {{3rd Dubna
  International Advanced School of Theoretical Physics}}}}\ (\bibinfo {year}
  {2005})\ \Eprint {http://arxiv.org/abs/hep-ph/0508242} {arXiv:hep-ph/0508242}
  \BibitemShut {NoStop}%
\end{thebibliography}%

\newpage

\onecolumngrid
\appendix*
\allowdisplaybreaks
\setcounter{secnumdepth}{2}


\section*{Supplemental material}

\subsection*{Canonical Basis}
\label{sec:bases}
In this section of the supplemental material, we provide the definition of our $\epsilon$-factorized basis for the non-planar six-propagator elliptic sector. 
In deriving it, we encounter  
several square roots, which should all be interpreted with 
the prescription $\dfS \to \dfS+i 0$, $\dfT \to \dfT+i 0$
 \begin{align}
 \label{eq:sqrts}
r_\dfS =  \sqrt{-\dfS} \sqrt{4 m^2-\dfS}\,,\quad 
 r_\dfT = \sqrt{-\dfT}\sqrt{4 m^2-\dfT}\,, 
 \quad
r_\dfU = \sqrt{-\dfS-\dfT} \sqrt{4 m^2-\dfS-\dfT}\,.
\end{align}

Introducing for convenience $\Psi_1= r_t \Psi_0$, the $\epsilon$-factorized 
basis for the sector corresponding to the right-hand graph of fig.~\ref{fig-non-planar-top} reads 
 \begin{align}
 \label{eq:pure-next-to-top}
 \rJ_{42}=&\, 
r_u \,( \rI_{1101111-10}-\rI_{110011100})\,,
  \nonumber\\
  \rJ_{43}=&\,
  r_\mathrm{s}
\,(
  \rI_{1101111-10}+ \rI_{11011110-1}-t\,  \rI_{110111100}
+  \rI_{011011100}-\rI_{111010100}+\frac{1}{D-4}\rI_{120010100}
-\frac{2}{D-4}\rI_{121000100}
 )\,,
 \nonumber\\  \rJ_{44}=&\,  
 r_\mathrm{t} \, (\rI_{1101111-10}-\rI_{11-1111100})+\left(\frac{(4m^2+s)t^2}{r_\mathrm{t}}+\frac{\rT_1}{\Psi_1}\right) \rI_{110111100}\,,
     \nonumber\\ 
       \rJ_{45}=&\left
      [ \frac{s+t-4m^2}{4m^2-s} t^2 \partial_t  +\frac{\rT_2}{\Psi_1}
      +\frac{1}{D-4}\frac{t}{2}
      \left(
      27m^2+2D(-3m^2+s)-8s+4m^2 \frac{t}{s-4m^2}
      \right)
      \right] \rI_{110111100}
                \nonumber\\ 
     +&\frac{s+t-4m^2}{2(4m^2-s)}\frac{1}{D-4}
     \bigg[ t \,\rI_{11-1111200}+
     t(s-2m^2)\rI_{110011200}-s(4m^2-s)\rI_{121010100}
     \bigg]
     \nonumber\\ 
     +&
     \frac{1}{4} t \left[
     2 \rI_{1101111-10}-2\rI_{11-1111100} +
     \rI_{010111100}-\rI_{111010100}+\frac{2}{D-4}\left(\rI_{012011100}-\rI_{210011100}\right)
     \right]
     \,,
      \nonumber\\ 
             \rJ_{46}=&\, 
             \bigg[
             \left(
            4t^2(s+t-4m^2)\rT_2+2t^2(s+t-4m^2)(-3m^2t+s^2+2s t-4 m^2s)\Psi_1
             \right)\frac{1}{s-m^2}\partial_t+ \frac{t^2(m^2+s)(3+\frac{4t}{s-m^2})}{4-D}\Psi_1
          \nonumber\\  
                       +&
                      \frac{96m^6 s t-24 m^4 s^2 t+108m^6 t^2-104 m^4s t^2+14 m^2 s^2 t^2 -43m^4 t^3+12 m^2s t^3-3 s^2 t^3}{2(t-4m^2)}\Psi_1      -\frac{\rT_1^2+4\rT_2^2}{2\Psi_1}    
                                \nonumber\\  
                      +&
                      \frac{2}{4-D}\left(
                      \frac{3 s t+4 t(t-3m^2)}{s-4m^2}\rT_2-2t^2(s+t-4m^2)(s+t)\partial_t \Psi_1
                      \right)+t(4m^2+s)\frac{r_t}{4m^2-t}\rT_1-4t(s-3m^2)\rT_2
                \nonumber\\     
           &  \bigg]\rI_{110111100}
             +\left[\frac{1}{D-4}\frac{2t^2(s+t-4m^2)(s+t)}{s-4m^2}\Psi_1 \right] \rI_{11-1111200}
             +\bigg[r_\mathrm{t} \rT_1+t^2(s+t)\Psi_1
             \bigg](\rI_{11-1111100}-\rI_{1101111-10})
               \nonumber\\ 
               +&
               t \bigg[-2\rT_2+t(t-m^2)\Psi_1\bigg]
\left(
\rI_{1101111-10}-\rI_{11-1111100}-\frac{1}{D-4}\frac{2(s+t-4m^2)}{s-4m^2}\rI_{11-1111200}
\right)
              \nonumber\\   
              +&
              \bigg[
              r_\mathrm{t} \rT_1+2 t \rT_2+(m^2+s)t^2 \Psi_1
              \bigg] 
              \left(
              \frac{1}{2}\rI_{111010100}-  \frac{1}{2}\rI_{010111100}+\frac{1}{D-4}\left(
              \rI_{210011100}-\rI_{012011100}
              \right)
              \right)
                     \nonumber\\ 
                     +&
                    \frac{1}{D-4}(s+t-4m^2) \bigg[
                     2\rT_2+t(m^2+s) \Psi_1
                     \bigg]
                     \left(
                     \frac{t(s-2m^2)}{s-4m^2}\rI_{110011200}-s\,\rI_{121010100}
                     \right)\,,
                 \nonumber\\        
      \rJ_{47}=& \,
      \frac{1}{\Psi_1}  \rI_{110111100}\,.
 \end{align}
 
Finally, we also provide the $\epsilon$-factorized basis for the top sector,
given in the left graph of fig.~\ref{fig-non-planar-top}. The three integrals
read
  \begin{align}
   \label{eq:pure-top}
  \rJ_{50}=&\,  r_\mathrm{s} r_u \,\rI_{1111111-10}\,,
     \nonumber\\ 
       \rJ_{51}=&\, r_\mathrm{s} 
       \left(
        \rI_{1111111-20}-t \,\rI_{1111111-10}-2 \rI_{1101111-10}+2 \rI_{110011100}
       \right)\,,
      \nonumber\\ 
      \rJ_{52}=&\, 
      s(4m^2-s) \left[
       \rI_{11111110-1}+\frac{1}{D-4}\left( \rI_{121010100}-2 \rI_{211100100}
       \right)
      \right]
    + \left(  \frac{2\rT_2}{\Psi_1}+t(2s-3m^2) \right)  \rI_{110111100}\,.
      \nonumber\\     
 \end{align}
 
\subsection*{UV and IR poles}
\label{sec:RC}
In this section, we collect useful formulas 
required to perform the renormalization of the UV poles of
the amplitude.
The renormalization constants  can be taken from~\cite{Broadhurst:1991fy,Melnikov:2000zc,Czakon:2007ej,Barnreuther:2013qvf,Bonciani:2021okt} and
to the required order they read
\begin{align}
 Z_e=& 1-\frac{\alpha_e(\mu)}{4\pi}
   \frac{\beta_0}{ \eps}
    +\left(\frac{\alpha_e(\mu)}{4\pi} \right)^2 \left(\frac{\beta_0^2}{\eps^2}-\frac{\beta_1}{2\eps}\right)+\mathcal{O}(\alpha_e(\mu)^3)\,,\\
Z_m=&1
    +\left(\frac{\alpha_e(\mu)}{4\pi}\right)  D_\eps \left(-\frac{3}{\eps}-4-8\eps-16\eps^2\right) +\mathcal{O}(\alpha_e(\mu)^2)\,,\\
   Z_2=&1+\left(\frac{\alpha_e(\mu)}{4\pi}\right)  D_\eps \left(-\frac{3}{\eps}-4-8\eps-16\eps^2\right) 
   + \left(\frac{\alpha_e(\mu)}{4\pi}\right)^{2} D_\eps^2\Bigg[  \left(\frac{1}{\eps}+\frac{947}{18}-5\pi^2\right)
    \nonumber \\
    +&  \bigg(\frac{9}{2\eps^2}+\frac{51}{4\eps}+\frac{433}{8}-13\pi^2 
    +16\pi^2\ln2-24\zeta_3\bigg) 
    + \beta_0\ln\left(\frac{\mu^2}{m^2}\right)\left(-\frac{3}{\eps}-4+\frac{3}{2}\ln\left(\frac{\mu^2}{m^2}\right)\right)\Bigg]+\mathcal{O}(\alpha_e(\mu)^3)\,,
\end{align}
where $D_{\eps}  = e^{\gamma_E\eps} \Gamma(1+\eps) \left(\mu^2/m^2\right)^{\eps}$, and the QED 
beta function coefficients are given by $\beta_0  =-\frac{4}{3}\,,   
\beta_1  = - 4$.
The  $\overline{\text{MS}}$ running coupling $\alpha_e(\mu)$ is related to the on-shell coupling $\alpha$ by~\cite{Grozin:2005yg}
\beq
\alpha_e(\mu)=\alpha\left[
1+\frac{4}{3}\frac{\alpha}{4 \pi}\ln \frac{\mu^2}{m^2}+\mathcal{O}\left(\frac{\alpha}{4 \pi}\right)^2
\right]\,.
\eeq

As described in the main text, IR singularities
exponentiate in QED in terms of the anomalous dimension $Z_1^{\textrm{IR}}$. For our process, its explicit form reads
\begin{align}
 Z_1^{\textrm{IR}}=&\frac{4(-2 m^2+s)}{\sqrt{-s}\sqrt{4m^2-s}}\ln\left(1-\frac{s}{2m^2}-\frac{1}{2}\sqrt{\frac{-s}{m^2}}\sqrt{4-\frac{s}{m^2}}\right)
 \nonumber\\
 +&\frac{4(-2 m^2+t)}{\sqrt{-t}\sqrt{4m^2-t}}\ln\left(1-\frac{t}{2m^2}-\frac{1}{2}\sqrt{\frac{-t}{m^2}}\sqrt{4-\frac{t}{m^2}}\right)
 \nonumber\\
 -&\frac{4(-2 m^2+u)}{\sqrt{-u}\sqrt{4m^2-u}}\ln\left(1-\frac{u}{2m^2}-\frac{1}{2}\sqrt{\frac{-u}{m^2}}\sqrt{4-\frac{u}{m^2}}\right)
  \nonumber\\
 -&4\,,
\end{align} 
whose analytic continuation to either Bhabha or Møller scattering
can be obtained by giving a positive imaginary part to the
Mandelstam variable that lies above the branch cut.

\end{document}